\documentclass[onecolumn, number, 5p]{elsarticle}

\usepackage[T1]{fontenc}
\usepackage[utf8]{inputenc}
\usepackage{amsmath}    
\usepackage{amssymb}    %
\usepackage{graphicx}   
\usepackage{verbatim}   
\usepackage{color}      
\usepackage{subfigure}  
\usepackage{hyperref}
\usepackage{multirow}
\usepackage{float}	
\allowdisplaybreaks

\begin{document}

\author[a]{F.~Maltoni}
\author[b]{M.~L.~Mangano}
\author[a]{I.~Tsinikos}
\author[c,d]{M.~Zaro}

\address[a]{Centre for Cosmology, Particle Physics and Phenomenology (CP3), Universit\'e Catholique de Louvain}
\address[b]{CERN, PH-TH, Geneva, Switzerland}
\address[c]{Sorbonne Universit\'es, UPMC Univ. Paris 06, UMR 7589, LPTHE, F-75005, Paris, France}
\address[d]{CNRS, UMR 7589, LPTHE, F-75005, Paris, France}

\title{Top-quark charge asymmetry and polarization in $t\bar{t}W^\pm$ production at the LHC} 

\date{\today}

\begin{abstract}
We study the charge asymmetry between the $t$ and $\bar t$ quarks at the LHC, when they are produced in association with  a $W$ boson. Though sizably reducing the cross section with respect to the inclusive production,  requiring a  $W$ boson in the final state has two important implications. First, at leading order in QCD,  $t \bar t W^{\pm}$ production can only occur via $q \bar q$ annihilation. As a result, the asymmetry between the $t$ and $\bar t$  generated at NLO in QCD is significantly larger than that of inclusive $t \bar t$ production, which is dominated by gluon fusion. Second, the top quarks tend to inherit the polarization of the initial-state quarks  as induced by the $W$-boson emission.  Hence,  the decay products of the top quarks display a sizable asymmetry already at the leading order in QCD.  We study the relevant  distributions and their uncertainties in the standard model,  compare them to those obtained in a simple axigluon model and discuss prospects for measurements at the LHC and beyond. 
\end{abstract}

\maketitle

\section{Introduction}
\label{sec:intro}

The charge asymmetry in $t \bar t$ production at $pp$ colliders is
defined by the quantity:
\begin{align}
&A_c^{t} = \frac{N(\Delta^t_\eta > 0) - N(\Delta^t_\eta < 0)}{N(\Delta^t_\eta > 0) + N(\Delta^t_\eta < 0)}\,, 
\label{eq:A_c}
\end{align}
\noindent
where $ \Delta^t_\eta = |\eta_t| - |\eta_{\bar t}|$. 
Quantum chromodynamics (QCD) predicts that radiative corrections to
the leading-order (LO) $t\bar{t}$ production process induce a
non-vanishing $A_c^t$, implying that top quarks are produced with a
rapidity distribution wider than anti-top quarks. At
next-to-leading order (NLO), this effect was first calculated
in Refs.~\cite{Kuhn:1998jr,Kuhn:1998kw}.    
The interest in this asymmetry stems from  measurements of the
corresponding forward-backward asymmetry ($A_{FB}$) performed in $p\bar p$
collisions at the Tevatron by the CDF and D$\emptyset$
Collaborations~\cite{Aaltonen:2012it,
Aaltonen:2011kc,Aaltonen:2014eva,Abazov:2011rq,Abazov:2014cca,Aaltonen:2014eva}.
These 
measurements point to a departure from the SM
predictions~\cite{Kuhn:1998jr,Kuhn:1998kw,Bowen:2005ap,Antunano:2007da,Almeida:2008ug,Hollik:2011ps,Manohar:2012rs,Bernreuther:2012sx},
giving rise to a large literature of possible interpretations based on
physics beyond the Standard Model (BSM) (for a recent review,
see~\cite{JA2014}).  

Unfortunately, the charge asymmetry predicted by the Standard Model
(SM) at the LHC is much smaller than $A_{FB}$ at the
Tevatron. The main reason is that both  $A_{FB}$ and $A_c$
are induced only by the fraction of $t\bar t$ final
states generated by $q\bar q$ collisions, which at the LHC represent
only $\sim 15\%$ of the total rate, compared to  $\sim 85\%$ at the
Tevatron.  The smallness of the effect makes it difficult at the LHC
to reach the sensitivity required to measure $A_c^t$, and to probe the
possible existence of BSM contributions, unless the BSM departures from
the SM prediction were rather large. 

The measurements of top charge asymmetry 
performed so far by the ATLAS and CMS Collaborations~\cite{Aad:2013cea,Chatrchyan:2012cxa,CMS:2013nfa,ATLAS:2012sla,Chatrchyan:2014yta},
have reached uncertainties at the level of $\delta A_c^t \sim 0.01$,
which is of the order of the SM value of $A_c^t$. For example, in their latest
publications relative to data taken at $\sqrt{s}=7$~TeV, 
ATLAS~\cite{Aad:2013cea} and CMS~\cite{Chatrchyan:2014yta}
report the following results~\footnote{The LHC measurements are
  reported in terms of rapidity differences $\Delta^t_y=\vert y_t\vert -
  \vert y_{\bar t}\vert$, and we shall refer to these asymmetries as
  $A_{c,y}$. Charge asymmetries based on rapidity differ by about
  10-20\% from those based on pseudorapidity,
  but have otherwise the same features. Here the leptonic asymmetries 
$A_{c}^{\ell}$ are  
defined by replacing in
Eq.~(\ref{eq:A_c}) $ \Delta^t_\eta $ with $
\Delta^{\ell}_{\eta} =  |\eta_{\ell^+}| - |\eta_{\ell^-}|$. In the
following we shall also consider  $A_c^{b}$, defined by  $
\Delta^b_\eta = |\eta_b| - |\eta_{\bar b}|$.}: 
\begin{align}
\mathrm{ATLAS:} &\; A_{c,y}^t=0.006\pm
  0.010_{\mathrm{stat+syst}} \, , \\
\mathrm{CMS:} &\; A_{c,y}^t=-0.010\pm 0.017_{\mathrm{stat}} \pm
    0.008_{\mathrm{syst}} \, , \\
\mathrm{CMS:} &\;\;\;\, A_{c}^\ell=0.009\pm 0.010_{\mathrm{stat}} \pm
    0.006_{\mathrm{syst}} \, .
\end{align}
A combination of the ATLAS and CMS results has also been performed by
the Top LHC Working Group~\cite{topwg}
\begin{equation}
A_{c,y}^t=0.005\pm0.007\pm0.006\,.
\end{equation}
The SM result used as a comparison in the experimental papers,
including both QCD and electroweak (EW) radiative corrections at the
one-loop level, is obtained from Ref.~\cite{Bernreuther:2012sx}: 
\begin{eqnarray}
A_{c,y}^t (7~\mathrm{TeV})&=&0.0123 \pm 0.0005 \, , \\
A_{c}^{\ell} (7~\mathrm{TeV})&=&0.0070 \pm 0.0003 \, . 
\end{eqnarray}
These values\footnote{The asymmetries for higher beam energies are
  determined in Ref.~\cite{Bernreuther:2012sx} to be 
$A_{c,y}^t (8~\mathrm{TeV})=0.0111 \pm 0.0004$ and
$A_{c,y}^t (14~\mathrm{TeV})=0.0067 \pm 0.0004$.} result from using the LO
total cross sections in the denominator of Eq.~(\ref{eq:A_c}). This is
justified by the fact that, at the one-loop level, the asymmetry is a
LO effect.  
Using the NLO total cross section, which is
$\sim 50\%$ larger than the LO one, the calculated asymmetries would be
reduced to $\sim 2/3$ of the above values. We believe that, in absence
of a complete NLO calculation of $A_c^t$, the difference between the
use of LO and NLO cross sections in the denominator of
Eq.~(\ref{eq:A_c}) should be included
in the estimate of the overall theoretical 
uncertainty. Should the true SM value of $A_c^t$ end up being closer
to the smaller values obtained using the NLO cross sections (e.g. $A_c^t\sim
0.004$ at $\sqrt{s}=14$~TeV), a robust
and accurate measurement will be a hard experimental challenge. 

Alternative observables are known to enhance the size of
the asymmetry. For example, Ref.~\cite{Bernreuther:2012sx} estimates
that the asymmetry can increase by a factor of 2-3 placing proper cuts
on the invariant mass of the $t\bar{t}$ system. The smaller rates
due to the extra cuts will
be compensated  by the much larger statistics to become
avilable at 13-14~TeV. But the theoretical systematics will, by and
large, remain correlated with those of the predictions for the 
underlying fully inclusive $A_c^t$. 

\begin{figure}[hf]
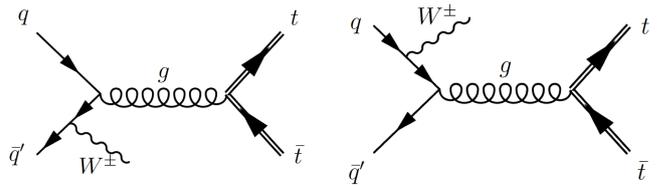

\centering
\subfigure{\includegraphics[width= 1.6in]{images/ttW_1}}
\hspace{.3cm}\subfigure{\includegraphics[width= 1.6in]{images/ttW_2}}
\caption{Feynman diagrams for the  $t \bar t W^{\pm}$ production at leading order in QCD.}
\label{fig:ttxW}
\end{figure}

In this work, we therefore consider an alternative production mechanism for top
quark pairs, which can provide a complementary handle for the
determination of the SM charge asymmetry, as well as an independent
probe of possible BSM sources of a deviation from the SM result.
The mechanism we propose is the production of a $t \bar t$ pair in
association with a $W$ boson (Fig.~\ref{fig:ttxW}).  This production
process is indeed quite peculiar. At the LO in QCD it can only occur
via a $q \bar q$ annihilation, and no contribution from gluons in the
initial states is possible. This is at variance with respect to $t
\bar t Z $ or $t \bar t \gamma $, where the vector boson can
also couple to the top quark in the subprocess $gg\to t\bar{t}$. As it
can be seen from Fig.~\ref{fig:ttxW}, $t \bar t W^\pm$ can be simply
thought of as the standard $q\bar{q}\to t\bar{t}$ LO diagram, with the
$W^\pm$ emitted from the initial state.  At the NLO, the $qg$ channels
can open up, yet the gluon-gluon fusion production is not accessible
until NNLO.  As in $q \bar q \to t \bar t$ the top and the anti-top
are produced symmetrically at LO and an asymmetry arises only starting
at NLO due to interference effects. As we will show in the following,
the absence of the symmetric gluon-gluon channel makes the resulting
asymmetry significantly larger than in $t\bar{t}$ production.

\noindent
The second key feature of $t \bar t W^\pm$ is that the emission of the
$W$ boson from the initial state acts as a polarizer for quark and
anti-quarks, effectively leading to the production of polarized top
and anti-top quarks. In other words, the $W$-boson emission makes the
production of a $t \bar t$ pair similar to that in polarized $e^+ e^-$
collisions
\cite{Ananthanarayan:2012ir,Brandenburg:1999ss,Groote:2010zf,Harlander:1994ac}.
As a result, the decay products of the top and anti-top display very
asymmetrical distributions in rapidity already at the leading
order. We shall call this the EW component of the 
asymmetry. In new physics scenarios, the emission of a $W$ boson might also
act as a discriminator of the chirality structure of new interactions,
such as that of an axigluon with light quarks, as already advocated in
different studies~\cite{Cvetic:1992qv,Guth:1991ab,Jezabek:1994zv}.

Results at the NLO and NLO+PS for the processes $t \bar t V \; (V =
W^\pm,Z)$ have appeared in the literature~\cite{Hirschi:2011pa,Campbell:2012dh,Garzelli:2012bn,Lazopoulos:2007bv,Lazopoulos:2008de,Kardos:2011na,Badger:2010mg}
yet no special attention has been given to asymmetries, whether EW or
QCD. The effect on
the asymmetry due to the emission of a photon has been recently
studied in Ref.~\cite{Aguilar-Saavedra:2014vta}. Measurements of total
rates are also becoming available from the LHC
experiments~\cite{Chatrchyan:2013qca}.  

The plan of this article is as follows. In Section~\ref{sec:NLO} we present the
predictions, at NLO in QCD, (with and without including parton shower and
hadronization effects) for $A_c^t$ in both $t \bar t$ and $t \bar t
W^\pm$ production, and, in the latter case, for the asymmetries of the
decay products $A_c^b$ and $A_c^{\ell}$.  In Section~\ref{sec:BSM}, we compare the
SM predictions to a simple benchmark model featuring an axigluon
compatible with the Tevatron $A_{FB}$ measurements, along the lines of
what done in Ref.~\cite{Falkowski:2012cu}, to illustrate the peculiar
discriminating power of $t \bar t W^\pm$. In the final section we
discuss the prospects at present and future colliders and present our
conclusions.  In \ref{sec:basics}, we review the main
features of the polarized $q \bar q$ annihilation into $t \bar t$,
highlighting the close similarity of angular distributions with those
predicted in $q\bar q \to t \bar t W^\pm$.

\begin{table}[t]
\renewcommand{\arraystretch}{1.5}
\begin{center}
\small
\begin{tabular}{ c | c | c c}
\hline\hline
$t \bar t$  & LO+PS &  NLO  & NLO+PS  \\ 
[1ex]
\hline
$ \sigma$(pb) & {$128.8^{+35\%}_{-24\%}{} $} & \multicolumn{2}{c}{$198^{+15\%}_{-14\%}{} $}  \\
[1ex]
\hline
 $A_c^{t}$ (\%) & $0.07 \pm 0.03$ & $0.61^{+0.10}_{-0.08}  $ & $0.72^{+0.14}_{-0.09} $ \\ 
[1ex]
\hline
\end{tabular}
\caption{Total cross sections and the asymmetry $A_c^{t}$ for $p p \rightarrow t
  \bar t$, calculated  at  NLO fixed order, LO+PS, and  NLO+PS  at $8 \; \textrm{TeV}$. 
  The quoted uncertainties are estimated with scale variations, except for LO+PS $A_c^{t}$ where they are from  MC statistics.
  For the NLO (+PS) $A_c^{t}$ MC uncertainties are less than 0.1 (absolute value in \%). } 
\label{table:tt}
\end{center}
\end{table}

\section{$t \bar t$ and $t \bar t W^{\pm}$ at NLO and NLO+PS}
\label{sec:NLO}

In order to study the top charge asymmetry at NLO for both $t \bar t$
and $ t \bar t W^{\pm} $, we employ {\sc MadGraph5\_aMC@NLO}, a
framework~\cite{Alwall:2014hca} which allows to automatically generate the
code needed to compute the cross section and any other observable for
these (and any other SM) processes at LO, NLO and NLO+PS. We present
results computed using the MSTW 2008 (N)LO PDF set
\cite{Martin:2009iq} with five massless flavors. The pole mass of the
top quark is set to 173 GeV and the $W$-boson mass to 80.41 GeV. The
renormalization and factorization scales are kept fixed and set to
$\mu_f = \mu_r = 2 m_t$, and the corresponding uncertainty is obtained
by varying the two scales independently in the interval $[ m_t , 4m_t
]$. PDF uncertainties are calculated following the Hessian recipe
given in Ref.~\cite{Martin:2009iq}. As we have found that they are negligible   
in the case of $A_c^t$ (at the level of 0.01 percent), we do not display 
them in the tables.  

We first show in Tab.~\ref{table:tt} the cross section and asymmetry
$A_c^{t}$  for $pp \rightarrow t \bar t$, computed at the LHC with a
center-of-mass energy $\sqrt{s}=8$~TeV. At the LO there is
no top-quark charge asymmetry, while including the parton shower
generates a small asymmetry, as shown in Ref.~\cite{Skands:2012mm}. 
At NLO a small asymmetry appears (less
than 1\%) both in the fixed order as well as in the NLO+PS
computation, the latter being slightly
larger~\cite{Skands:2012mm}. Not surprisingly, a rather strong scale
dependence affects the asymmetry predictions, these being {\it de
  facto} LO quantities. Here and below, we shall always use the NLO
cross sections in the denominators of the asymmetries, leading to a
possible underestimate of the real asymmetry. As stated above, we
believe that the difference between using LO and NLO cross sections in
the denominators should anyway be considered as an additional component of the
theoretical systematics (though, this is not accounted for in the scale
uncertainties quoted throughout the paper). 

\begin{figure}[h!]
\centering
\subfigure{\includegraphics[width= 0.5\textwidth]{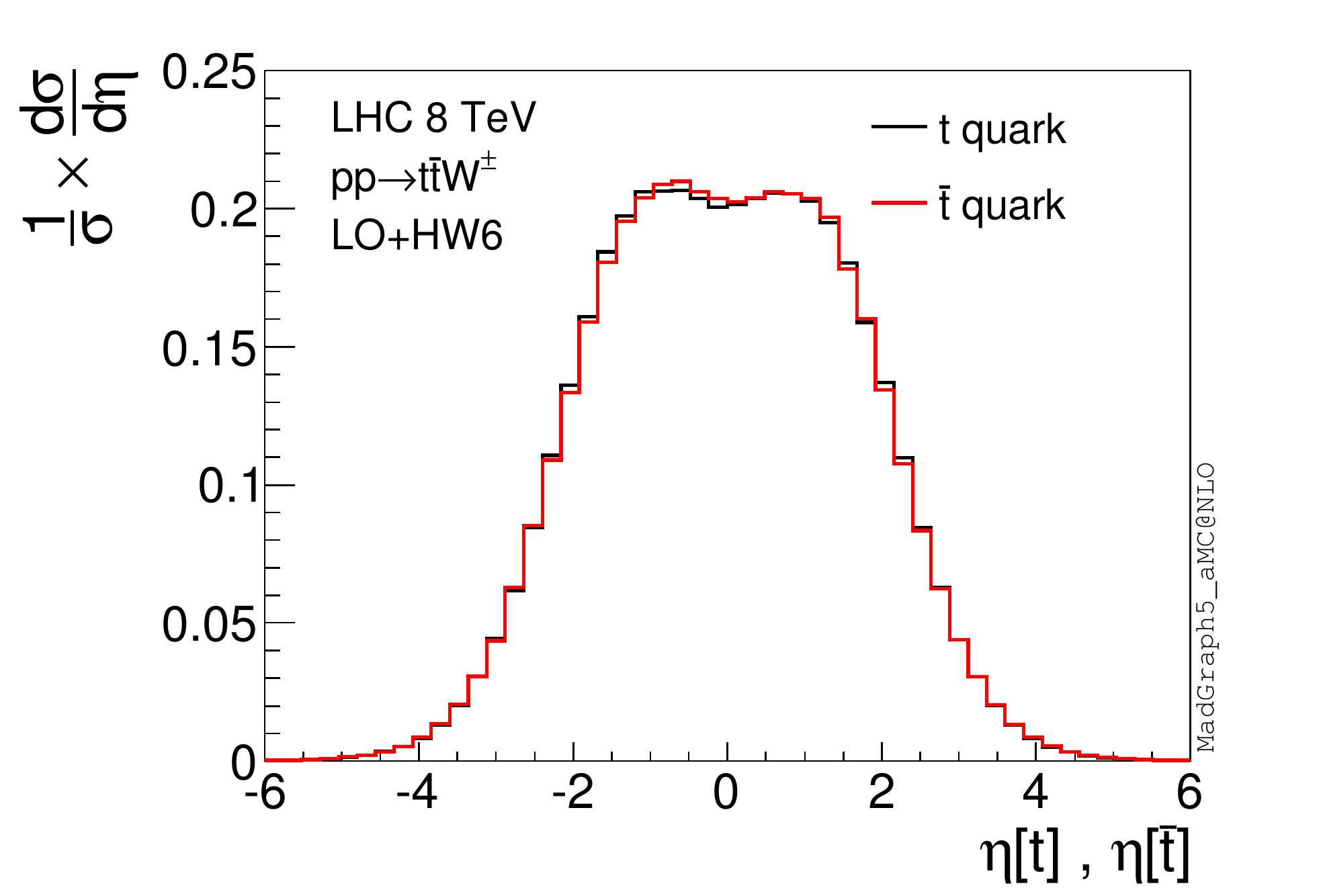}}
\subfigure{\includegraphics[width= 0.5\textwidth]{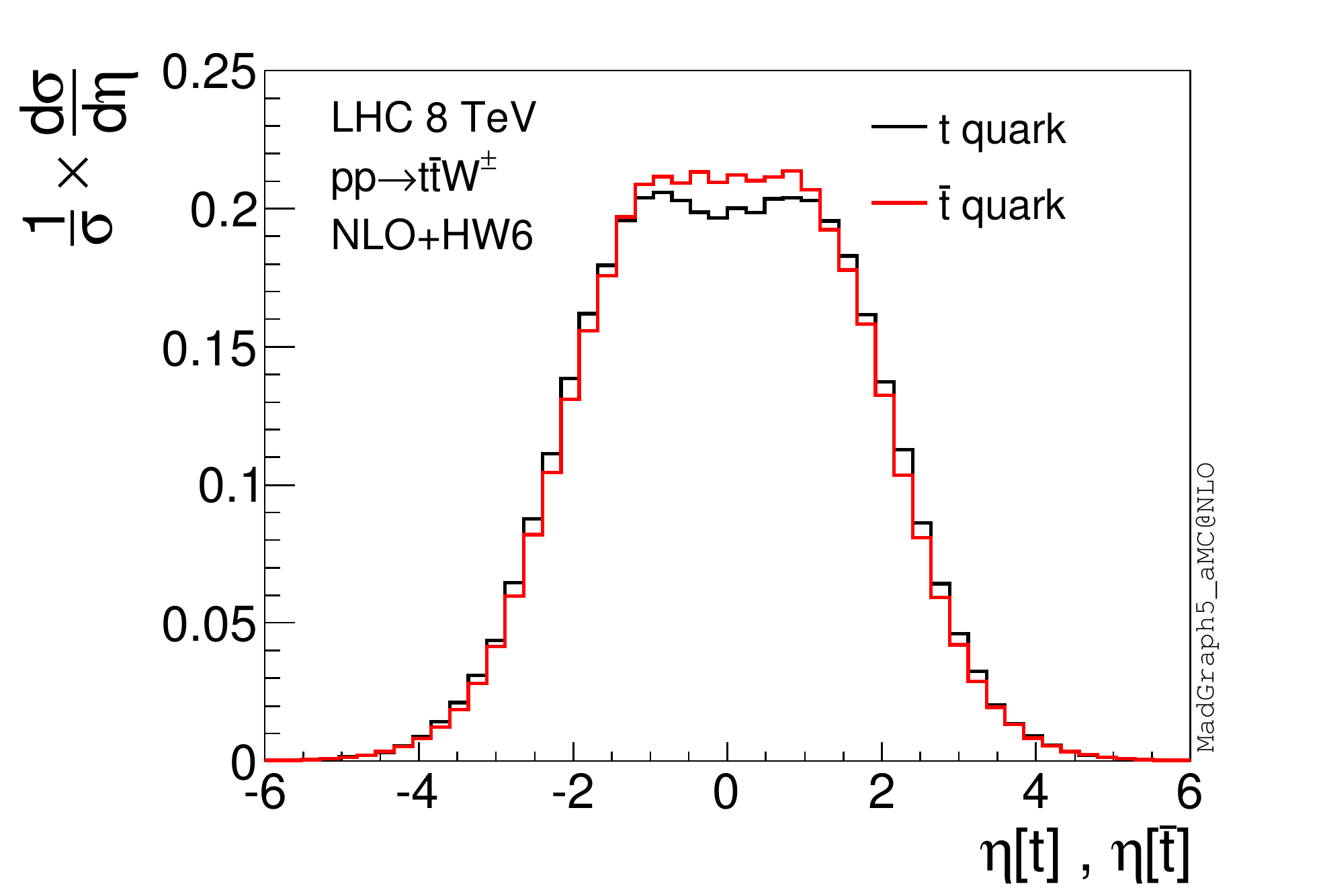}}
\caption{Comparison of the $\eta$ distributions of the $t , \bar t$
  quarks at the (N)LO+PS level for the $pp \rightarrow t \bar t
  W^{\pm}$ channel.} 
\label{fig:Comp_ttbar}
\end{figure}

\begin{table*}[t]
\renewcommand{\arraystretch}{1.5}
\begin{center}
\small
\begin{tabular}{ l | c | c c c  }
\hline
\hline
 & Order  & $ t \bar t W^{\pm}$ & $ t \bar t W^{+}$ & $ t \bar t W^{-}$  \\ [1ex]
 \hline
\multirow{2}{*}{$\sigma$(fb)} & LO & $140.5^{+27\%}_{-20\%}\;$ &
 $98.3^{+27\%}_{-20\%}\;$ &
 $42.2^{+27\%}_{-20\%}\;$ \\ 
 & NLO & $210^{+11\%}_{-11\%}$ 
       & $146^{+11\%}_{-11\%}$ 
       & $63.6^{+11\%}_{-11\%}$\\
\hline
\multirow{2}{*}{$A_c^{t}$ (\%)} 
 & NLO &  $ 2.49^{+0.75}_{-0.34}$ & $ 2.73^{+0.74}_{-0.42}$ & $ 2.03^{+0.81}_{-0.19}$ \\
 & NLO+PS& $2.37^{+0.56}_{-0.38} $  & $2.51^{+0.62}_{-0.42} $ &  $1.90^{+0.51}_{-0.35}$ \\ 
\hline
\end{tabular}
\caption{ Total cross sections (LO and NLO) and the asymmetry $A_c^{t}$ (NLO and NLO+PS)
  for $p p \rightarrow t  \bar t W^{\pm}$ at $8 \; \textrm{TeV}$. The quoted uncertainties 
  are estimated with scale variations. For the asymmetries MC uncertainties are less than 0.1 (absolute value in \%).} 
\label{table:fixed}
\end{center}
\end{table*}

\begin{figure}[t]
\centering
\includegraphics[width=0.5\textwidth]{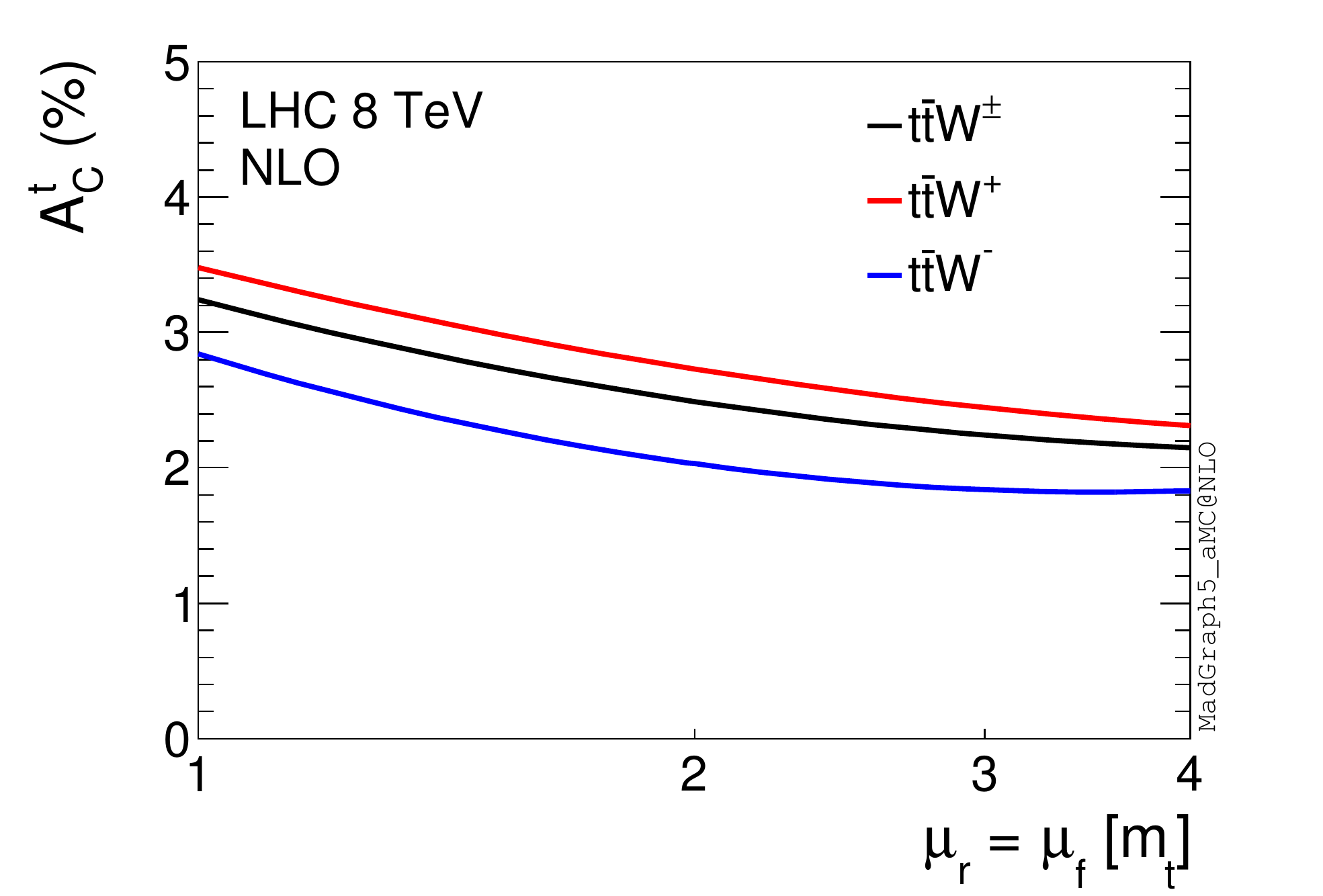}
\caption{$A_c^t$ asymmetry at fixed NLO.}
\label{fig:AsymmetryFO}
\end{figure}

We now turn to the corresponding results for $t \bar t W^{\pm} $,
which are shown in Tab.~\ref{table:fixed}. As in the previous case,
$A^t_c$ vanishes at
the LO, but at NLO we obtain
$A_c^t\approx2-3\%$,  a considerably larger value than in the $t\bar
t$ inclusive production. The effect of the asymmetry can be visualized
by superimposing the pseudorapidity of the $t$ and  $\bar t$ quarks,
as shown in  Fig.~\ref{fig:Comp_ttbar}.  
At LO the two distributions are not distinguishable, while at NLO the
asymmetry is manifest: the anti-top quark tends to be more central,
whereas the top quark has a broader spectrum, with a dip at $\eta=0$. 
Again, the scale dependence of the
asymmetry is quite large, consistently with the fact that NLO
corrections only provide its LO contribution. The
scale dependence of the asymmetry is shown in
Fig.~\ref{fig:AsymmetryFO}, varying  the renormalization and
factorization scales together. 

It is also worth to briefly comment on the fact that the asymmetry is
larger for $t \bar t W^+$  than for $ t \bar t W^-$. This can be
understood using an argument based on PDF's: the main subprocesses in
these two channels are $u \bar d \rightarrow t \bar t W^+$ and $d \bar
u \rightarrow t \bar t W^-$,  respectively. The longitudinal momenta
of the initial partons are on average  $p_u > p_d > p_{\bar u} \approx
p_{\bar d}$. In both cases the momentum of the $t \; (\bar t)$ quark
is connected to the momentum of the $q \; (\bar q)$. The large
longitudinal momentum transferred to the $t$ quark from the initial
$u$ quark ($t \bar t W^+$) increases the corresponding $|\eta_t|$
value. As a result the asymmetry $A_c^t$ is enhanced compared to the $ t \bar t W^-$ final
state.  

As a next step, we consider the case of a NLO+PS
simulation, obtained by matching the NLO calculation to {\sc Herwig6} \cite{Corcella:2000bw} via the {\sc MC@NLO} method~\cite{Frixione:2002ik}. We show the corresponding results in the third line of
Tab.~\ref{table:fixed}. The asymmetry at LO+PS (not shown in the
table) remains zero within uncertainties.  At the  NLO+PS level  a
small decrease compared to fixed NLO is found.  

Finally,  we analyze the results obtained including the decays of the
top quarks and the $W$-boson. In order to keep spin correlations
intact for the final lepton and $b$, $\bar b$ distributions, {\sc
  Madspin}~\cite{Artoisenet:2012st} is employed. In so doing
parton-level events are decayed using the full tree-level matrix
element  
$2 \to 8$ for the Born-like contributions and $2 \to 9$ for those
involving extra radiation, before they are passed to {\sc Herwig6}.

\begin{table*}[ht]
\renewcommand{\arraystretch}{1.5}
\begin{center}
\small
\begin{tabular}{ c | c |  c c c }
\hline\hline
 & Order &  $ t \bar t W^{\pm}$  & $ t \bar t W^+$  & $ t \bar t W^-$ \\ 
[1ex] 
\hline 
\multirow{2}{*}{$A_c^{b}$ (\%)} & LO+PS  &  $7.46^{+0.04}_{-0.05}$ &  $8.04^{+0.05}_{-0.06}$  &  $5.67^{+0.01}_{-0.01}$  \\[1ex] 
\cline{2-5}
 & NLO+PS &  $8.50^{+0.15}_{-0.10} $  & $9.39^{+0.15}_{-0.10} $ & $6.85^{+0.14}_{-0.11} $ \\ 
[1ex] 
\hline
\multirow{2}{*}{$A_c^{\ell}$ (\%)} & LO+PS & $-17.10^{-0.09}_{+0.11}$  &  $-18.65^{-0.12}_{+0.14}$ &  $-13.53^{-0.01}_{+0.03}$  \\  
[1ex] 
\cline{2-5}
 & NLO+PS & $-14.83^{-0.65}_{+0.95} $  & $-16.23^{-0.72}_{+1.04}  $ & $-11.97^{-0.50}_{+0.75} $ \\ 
[1ex]  
\hline
\end{tabular}
\caption{Asymmetries $A_c^{b,\ell}$, calculated at LO+PS and NLO+PS  level,  for $p p \rightarrow t \bar t W^{\pm}$ at $8 \; \textrm{TeV}$. The quoted uncertainties 
  are estimated with scale variations. Figures in the table have around 0.1\% of statistical uncertainty.}
\label{table:Madspin}
\end{center}
\end{table*}

At this exploratory stage, we use the MC truth in order to correctly
identify leptons and $b$-jets coming from the top and anti-top quark
decays, without considering issues related with the top quark
reconstruction. 
Furthermore we ask that the leptons coming from top (anti) quark
decays are positrons (electrons), while the extra $W$ bosons decay
into muons, requiring the following decay chains:\\ 

\textbullet \; $t \rightarrow b W^+ \rightarrow b e^+ \nu_e$\;\;\;\;\;
\textbullet \; $\bar t \rightarrow \bar b W^- \rightarrow \bar b e^-
\bar \nu_e$ \\ 

 \textbullet \; $W^- \rightarrow \mu^- \bar
 \nu_\mu$\;\;\;\;\;\;\;\;\;\;\;\;\;\; \textbullet \; $W^+ \rightarrow
 \mu^+ \nu_\mu$\,. \\\\ 
\noindent
We present the asymmetries $A_c^b$  and $A_c^{\ell}$ in
Tab.~\ref{table:Madspin}. The former is computed by reconstructing the
$b$-jets in the event which come from the top and anti-top quarks.  We
cluster hadrons into jets using the $k_T$ algorithm as implemented in
{\sc FastJet} \cite{Cacciari:2011ma}, with $R=0.7$, $p_T > 20 \;
\rm{GeV}$ and $|\eta | < 4.5$. Smaller values of the $R$
parameter have been checked not to alter significantly the results. For 
the computation of $A_c^b$, events that do not feature two $b$-jets coming from 
the top quarks have been discarded. 

\begin{figure}[hf]
\centering
\subfigure{\includegraphics[width= 0.5\textwidth]{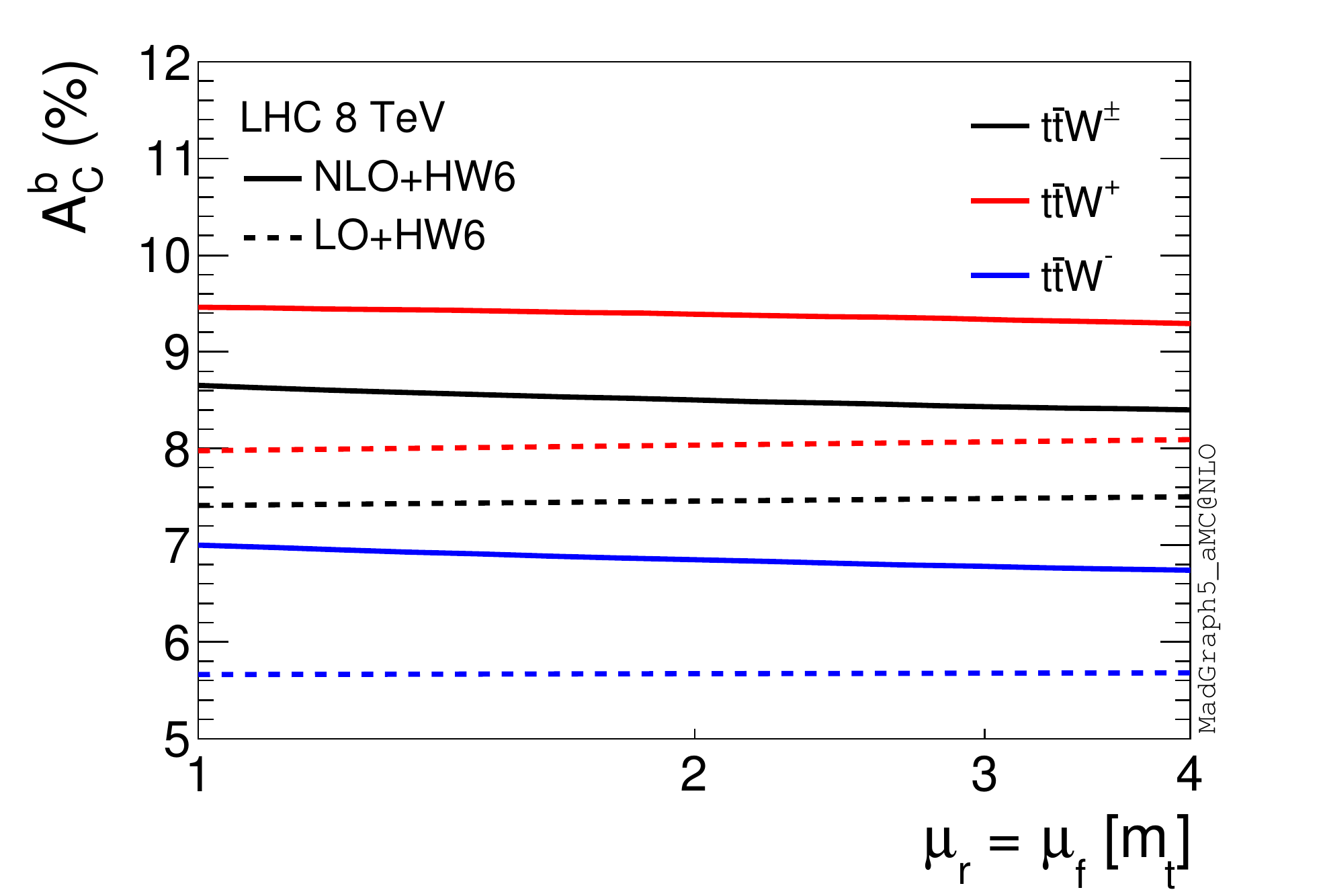}}
\subfigure{\includegraphics[width= 0.5\textwidth]{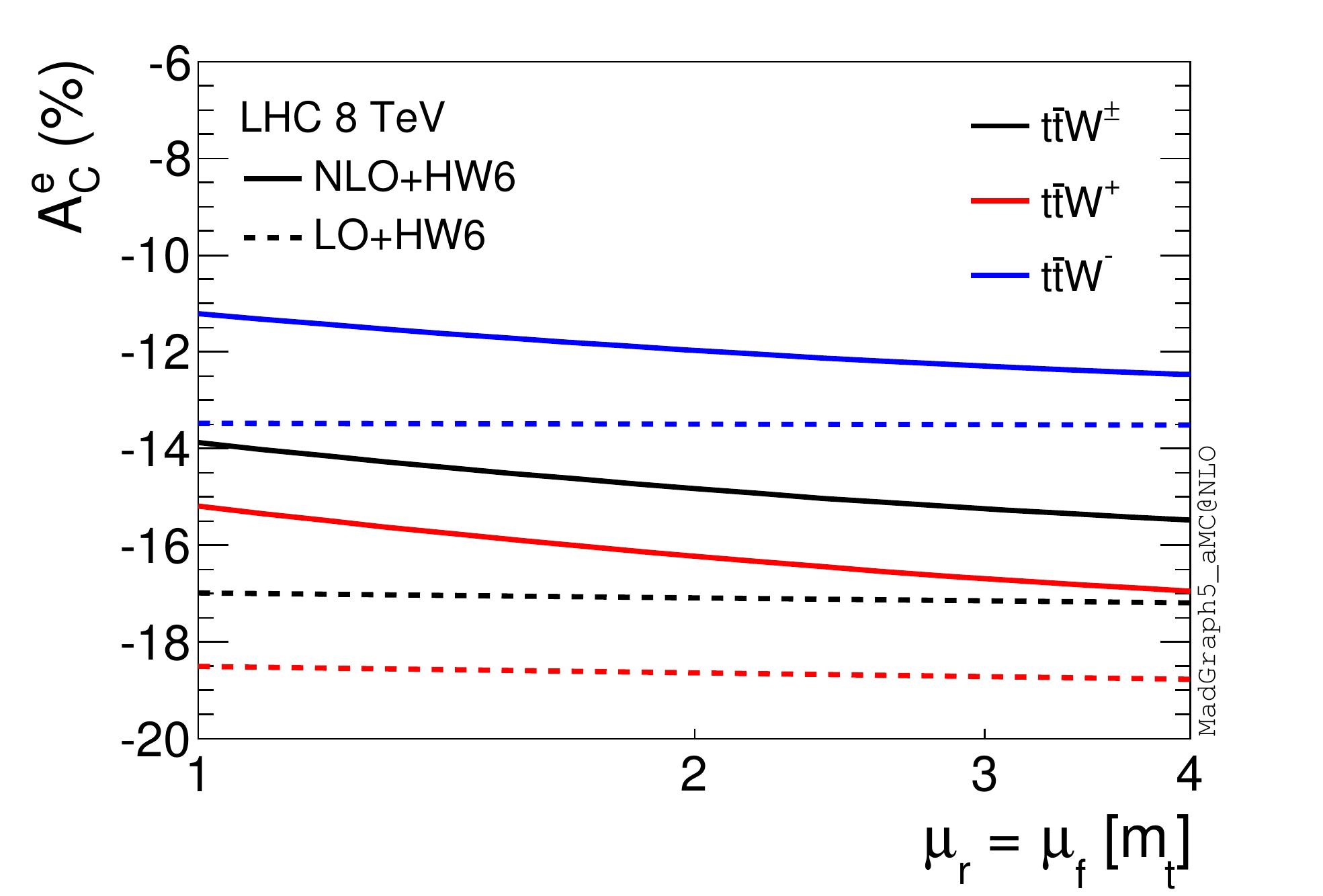}}\\
\caption{Comparison of the $A_c^b , A_c^{\ell}$ asymmetries between  LO+PS and NLO+PS. For all the three channels the dashed line (solid) is the LO+PS (NLO+PS).}
\label{fig:Asymmetry_b_e}
\end{figure}
\noindent

Two observations on the effects of NLO corrections can be made. The
first is that for both $A_c^\ell$ and $A_c^b$ NLO corrections tend to
shift the EW asymmetries towards positive values, an effect which is
consistent with $A_c^t$ being positive at the NLO. It is not possible
to exactly factorize the EW and QCD components of $A_c^\ell$ and $A_c^b$,
but one can estimate the intrinsic QCD part by suppressing the polarization
correlations in the decays, thus removing the EW contribution. 
In this case, we obtain $A_c^\ell = 1.79$ and
$A_c^b=2.0$, comparable to $A_c^t=2.37$.  

The second observation is that the scale dependence of these asymmetries is very
small at the LO, while it becomes larger at the NLO, as it can be seen
in Fig.~\ref{fig:Asymmetry_b_e}. This is due to the fact that 
the  asymmetry at LO is purely EW in origin, and it therefore rather
stable against scale variations, while the asymmetry at NLO includes
the QCD effects, and is directly affected by the scale dependence.

\section{BSM : the axigluon model}
\label{sec:BSM}

\begin{figure}[hf]
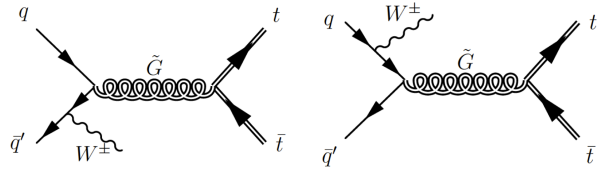

\centering
\subfigure{\includegraphics[width= 1.5in]{images/ttW_Ax_1}}\hspace{0.1in}
\subfigure{\includegraphics[width= 1.5in]{images/ttW_Ax_2}}
\caption{Leading order Feynman diagrams for the $pp \rightarrow t \bar t W^{\pm}$ process via an s-channel axigluon.}
\label{fig:ttxW_Ax}
\end{figure}
\noindent

The Tevatron experiments (CDF, D$\emptyset$) have measured the forward-backward asymmetry, which is defined in a similar way to the peripheral-central asymmetry used for the LHC, {\it i.e.}, 
\begin{equation}
A_{t \bar t} = \frac{N(\Delta^{t \bar t}_\eta > 0) - N(\Delta^{t \bar t}_\eta < 0)}{N(\Delta^{t \bar t}_\eta > 0) + N(\Delta^{t \bar t}_\eta < 0)} \; , \; \Delta^{t \bar t}_\eta = \eta_t - \eta_{\bar t}\,.
\end{equation}

The central values of the measurements from the two collaborations \cite{Aaltonen:2012it,Abazov:2011rq} are larger than the  SM~\cite{Bernreuther:2012sx,Hollik:2011ps}, 
\begin{align*}
\text{CDF: } A_{t \bar t} = 16.4 \pm 4.7 &\; \% \, , \; \text{D$\emptyset$: } A_{t \bar t} = 19.6 \pm 6.5 \; \% \, , \\
 \text{SM: } A_{t \bar t} &= 8.8 \pm 0.6 \; \% \,.
\end{align*}
A simple toy model that is often used to describe the enhancement of
the forward-backward asymmetry at the Tevatron features a massive
color octet vector boson (axigluon, ${\tilde G}$)
\cite{Ferrario:2009bz,Frampton:2009rk} that in general can couple in a different way
to light and heavy quarks. The enhancement is a result of interference
between the LO SM (Fig. \ref{fig:ttxW}) and the BSM amplitude
(Fig.~\ref{fig:ttxW_Ax}). Studies have already been performed in order
to calculate the forward-backward (Tevatron) as well as the
central-peripheral (LHC) asymmetry predicted by this model for the $t
\bar t$ channel \cite{Falkowski:2012cu,Carmona:2014gra}. 

We now study how an axigluon would manifest itself in a $t \bar t
W^\pm$ final state. To this aim, and to keep this part as simple as
possible, we restrict our BSM predictions to the LO. In general, the
axigluon couplings to the quarks are related to the strong coupling
$g_s$ and can either be universal (same couplings between light and
top quarks) or non-universal. The parameter space of the model
provides the freedom to choose a light or a heavy axigluon.  The term
in the Lagrangian that describes the coupling of the axigluon to
quarks is  
\begin{eqnarray}
\mathcal L &=& \sum_i \tilde G^{\mu,a} \left[ \right. g_L^i \bar q_i T^a \gamma_\mu (1- \gamma^5) q_i \nonumber \\
&+& g_R^i \bar q_i T^a \gamma_\mu (1+ \gamma^5) q_i \left. \right] \; , \; i = u, d, t\,.
\end{eqnarray}
\noindent
By choosing $g_L^i = 0 , g_R^i \neq 0$, a pure right-handed coupling
is possible. However, as argued in \ref{sec:basics}, in $t\bar t
W^\pm$ the initial quark line can only  be left-handed, leading to
vanishing amplitudes for right-handed axigluons. This brings the first
very important difference with respect to $t \bar t$ production whose
total rates are not sensitive to the relative amount $L$ and $R$
chiralities of the couplings of the axigluon. This is just an example
of a more general point: comparing  asymmetries in  $t \bar t$ and
$t\bar t W^\pm$ (and also other associated productions such as with
$Z$ and $\gamma$) could provide further key information on the new
physics interactions.  To illustrate this in the case of the axigluon
in a quantitative way, we consider four scenarios, two (left, axial)
for a light axigluon and two for a heavy one. The light axigluon is
chosen to have universal couplings and a fixed width $\Gamma_{\tilde
  G} = 50$  GeV, while the  heavy axigluon is chosen  to have
non-universal couplings and of opposite sign between the light and top
quark couplings. The model used for the light axigluon is  available
from the {\sc FeynRules} model database \cite{Alloul:2013bka,topBSM}
though it has been slightly modified in order to include non-universal
couplings for the heavy axigluon. For the light axigluon ($m_{\tilde
  G} = 200 \; {\rm GeV}, \Gamma_{\tilde G} = 50 \;  {\rm GeV}$) the
scenarios considered are:\\ 

Left-handed (I) : $g_L^u = g_L^d = 0.5 g_s\, , \quad g_R^u = g_R^d = 0\, ,$\\

Axial (II) : $g_L^u = g_L^d = -0.4 g_s\, , \quad g_R^u = g_R^d = 0.4 g_s\, .$\\[5pt]
\noindent
For the heavy axigluon ($m_{\tilde G} = 2$  TeV) the decay width is calculated internally \cite{Alwall:2014bza} considering the decays of the axigluon to quarks 
(and using $\alpha_S(m_{\tilde G})$) the scenarios are:\\

Left-handed (III) : \\[10pt]
\hspace*{.5cm} $g_L^u = g_L^d = -0.8 g_s \, , \quad g_R^u = g_R^d = 0 \, ,$ \\[5pt]
\hspace*{.5cm} $g_L^t = 6 g_s \, , \quad g_R^t = 0 \, , \quad  \Gamma_{\tilde G} = 1123 \; {\rm GeV}\, .$\\

Axial (IV) : \\[10pt]
\hspace*{.5cm} $g_L^u = g_L^d = 0.6 g_s \, , \quad g_R^u = -0.6 g_s\, , \quad g_R^d = 0 \, ,$\\[5pt]
\hspace*{.5cm} $g_L^t = -4 g_s \, , \quad g_R^t = 4 g_s \, , \quad  \Gamma_{\tilde G} = 742 \; {\rm GeV} \,.$\\[5pt]

\noindent
In order to calculate the asymmetries at the best of our knowledge we combine additively the NLO prediction  for the SM to the BSM one at LO, {\it i.e.},  
\begin{equation*}
\sigma_{\rm tot} \equiv \sigma^{\rm SM}_{\rm NLO} + \sigma^{\rm BSM}_{\rm LO}  ,
\end{equation*}
\noindent
where 
\begin{equation*}
\sigma^{\rm BSM}_{\rm LO} =  |A_{\rm {\tilde G}} + A_{\rm SM} |^2 -  |A_{\rm SM}|^2   ,
\end{equation*}
{\it i.e.},  the contribution of the diagram featuring the axigluon exchange squared as well as the interference with the SM amplitude. For consistency, we employ NLO PDF's in both SM and BSM terms. 

\begin{figure}[h]
\centering
\subfigure{\includegraphics[width= 0.5\textwidth]{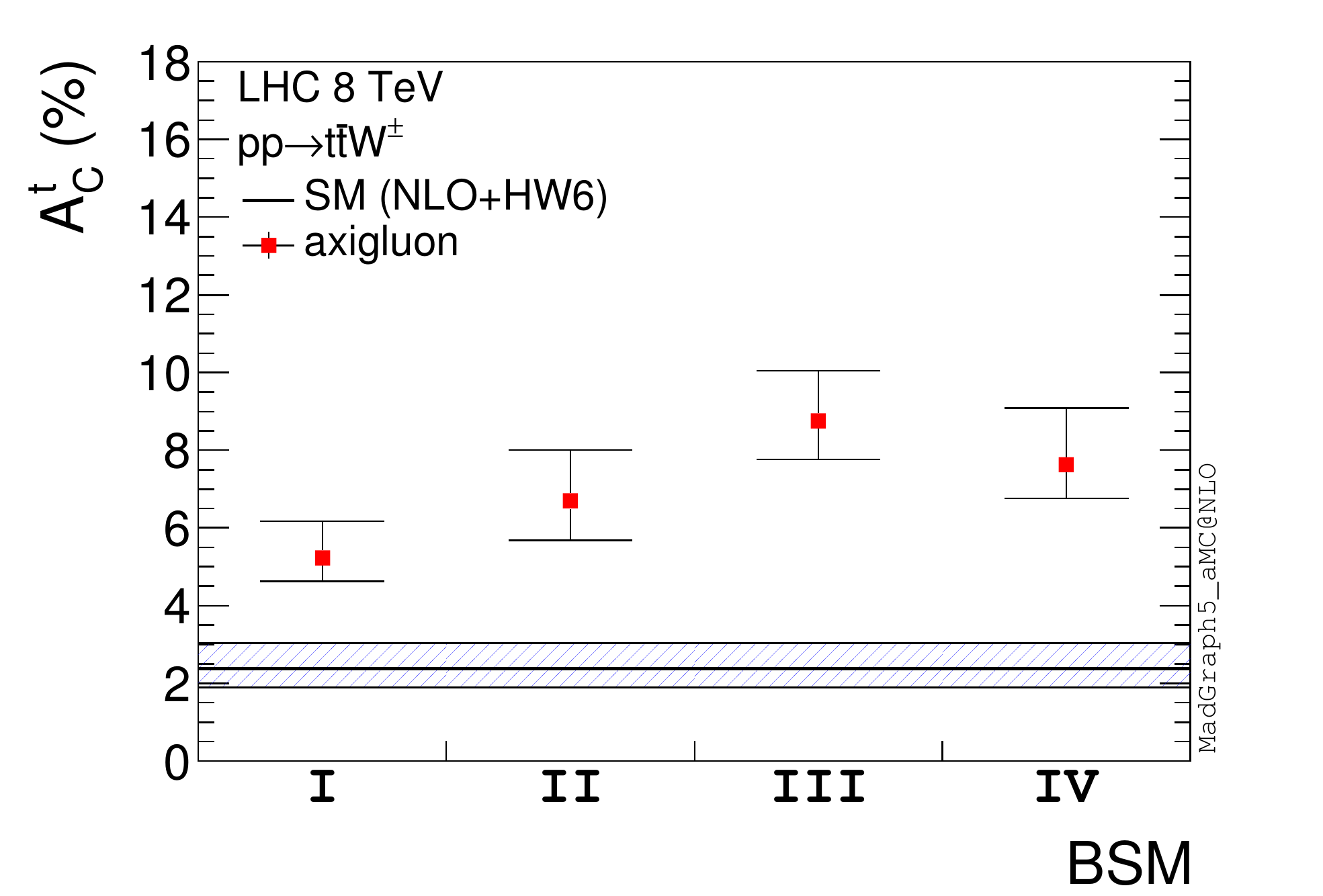}}\\
\subfigure{\includegraphics[width= 0.5\textwidth]{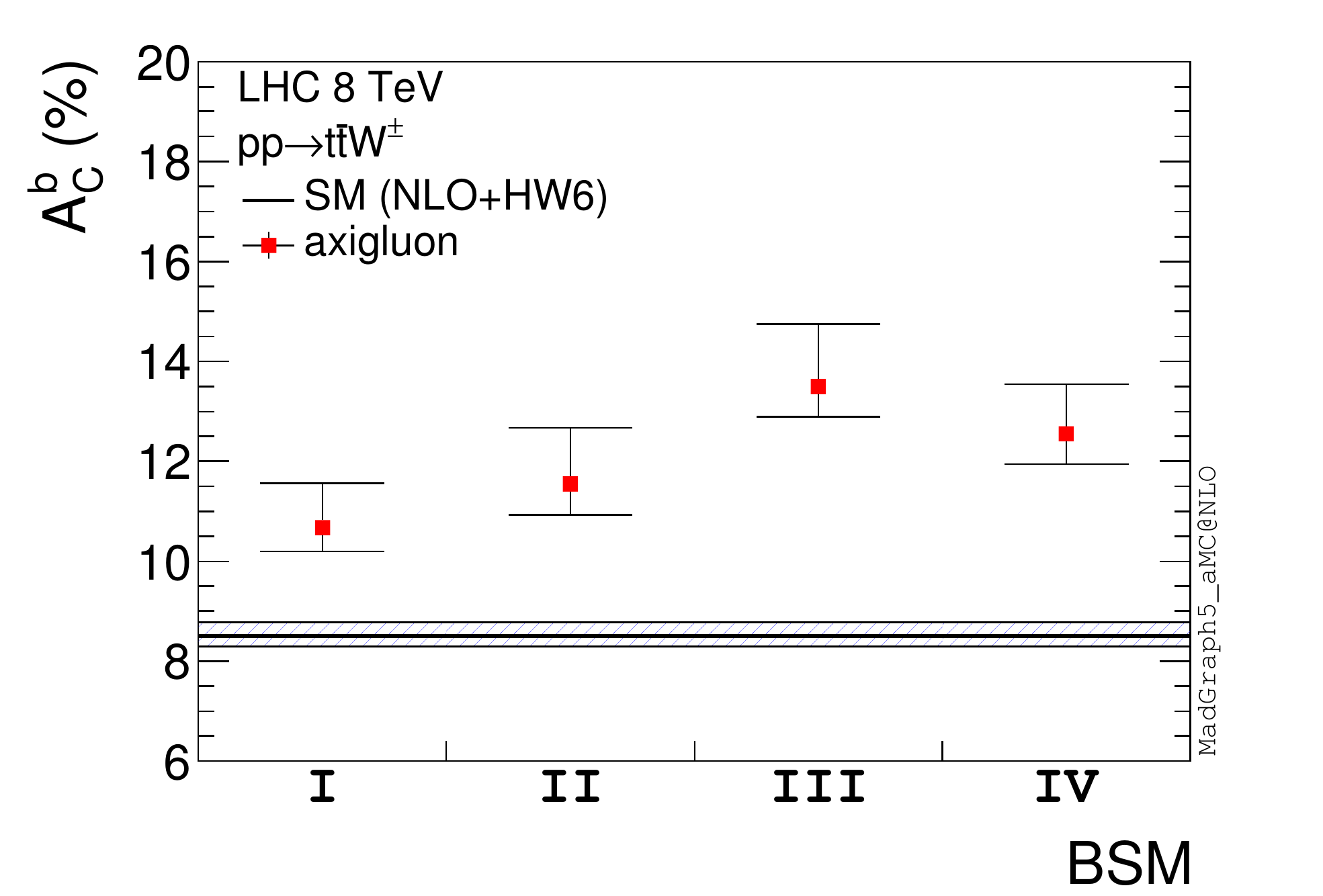}}\\
\subfigure{\includegraphics[width= 0.5\textwidth]{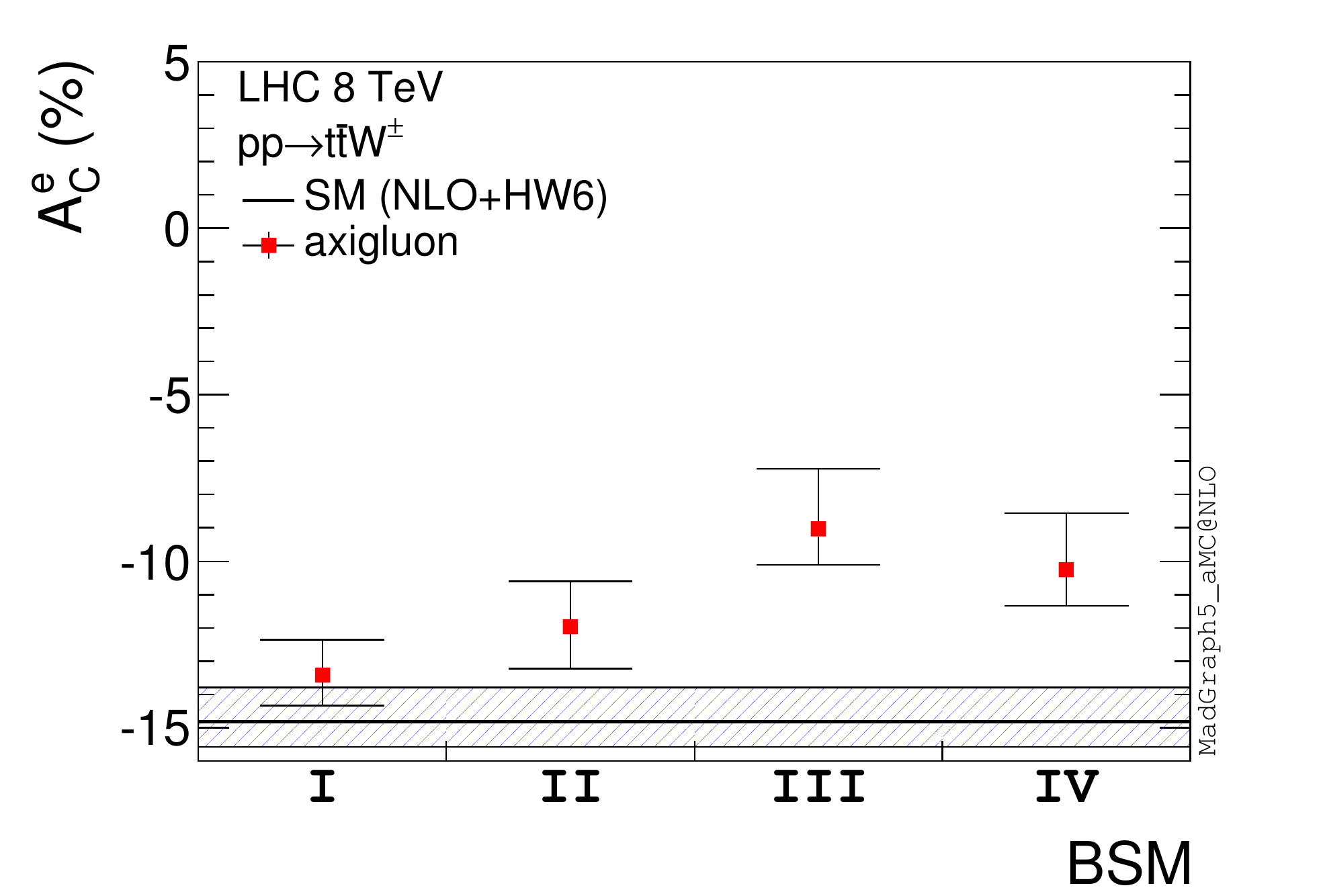}}\\
\caption{Comparison between asymmetries predicted in the axigluon (NLO+BSM, $I-IV$)  scenarios and the NLO SM prediction at 8 TeV, including scale  uncertainties.}
\label{fig:Total_Effect_8}
\end{figure}

\begin{figure}[h]
\centering
\subfigure{\includegraphics[width= 0.5\textwidth]{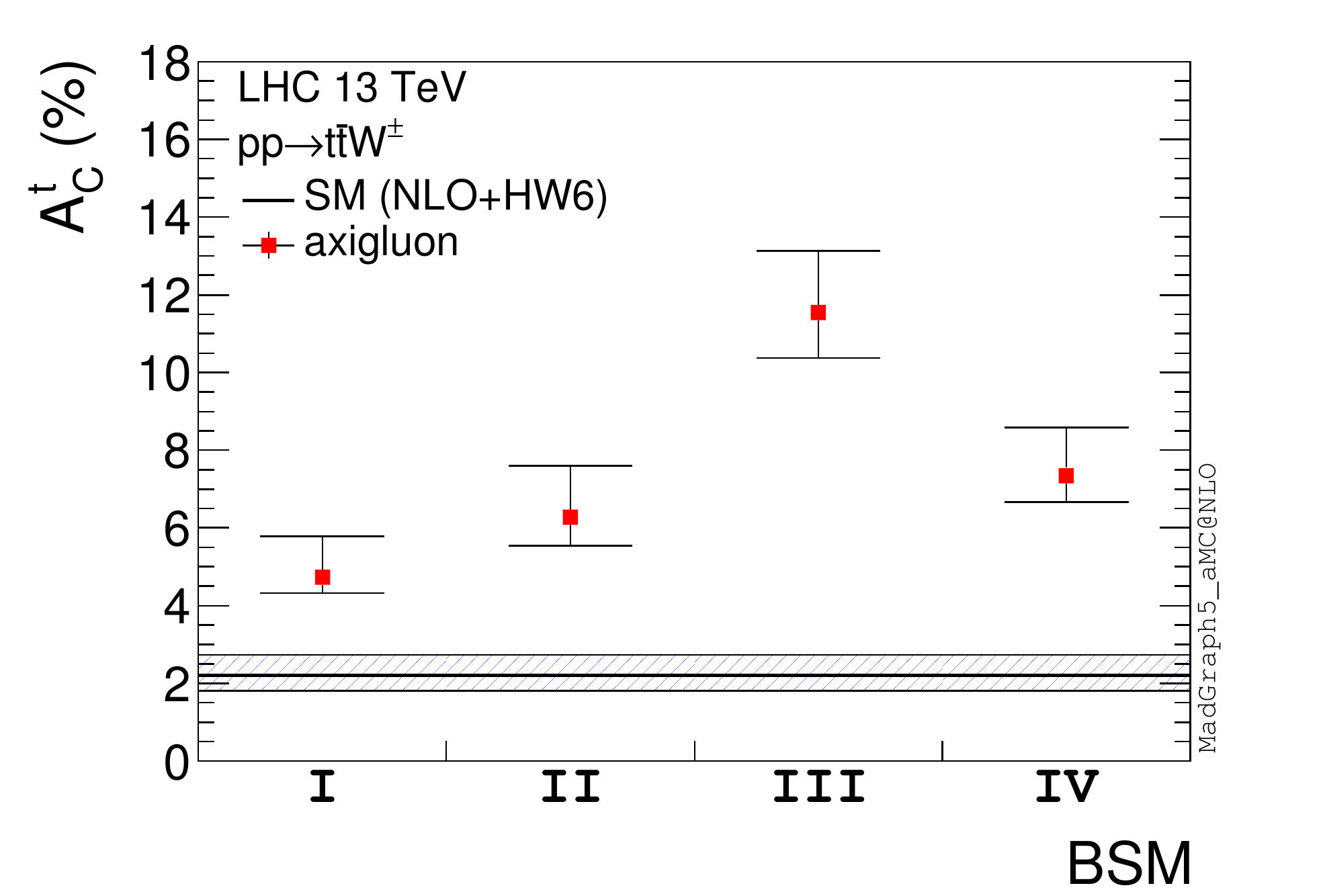}}\\
\subfigure{\includegraphics[width= 0.5\textwidth]{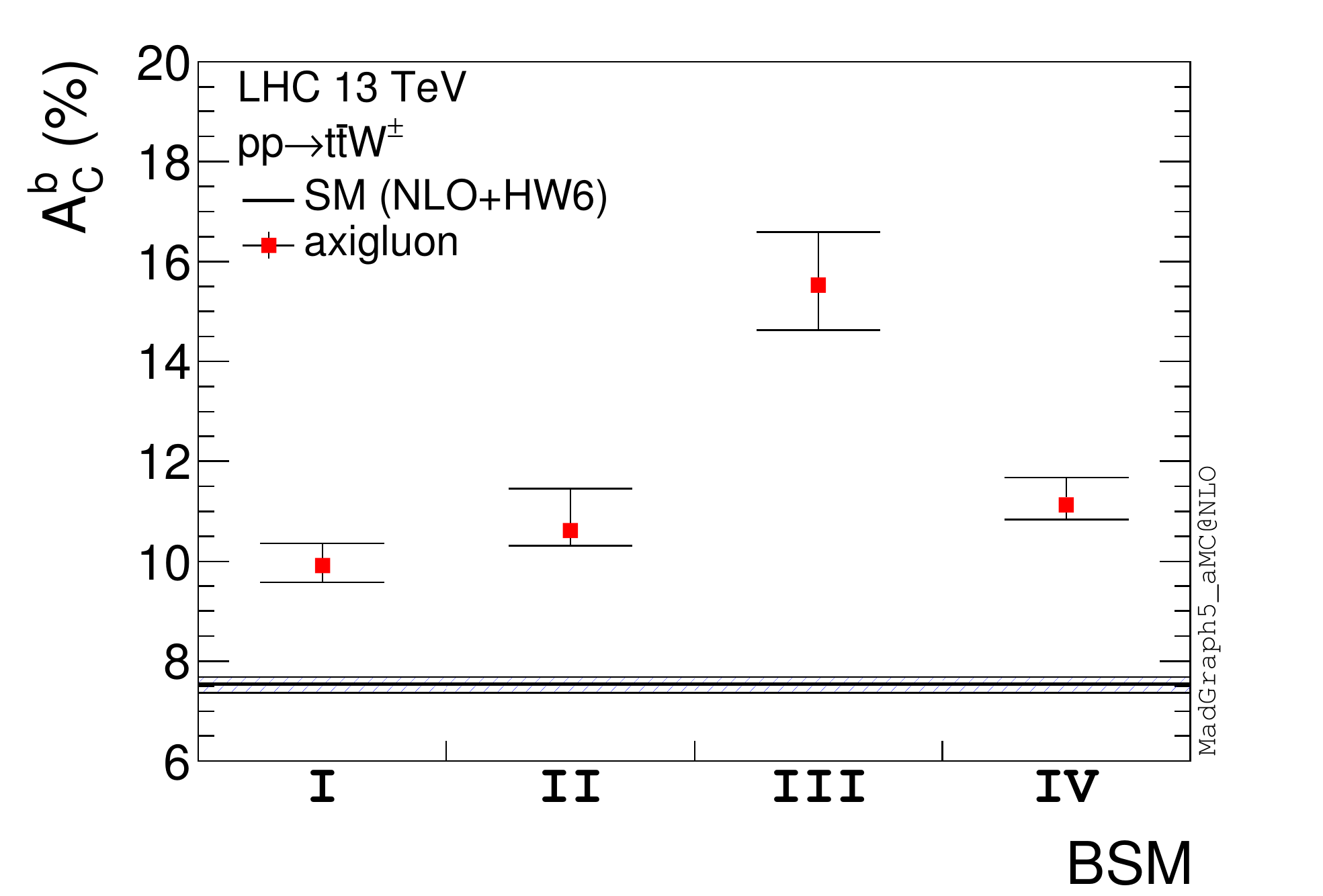}}\\
\subfigure{\includegraphics[width= 0.5\textwidth]{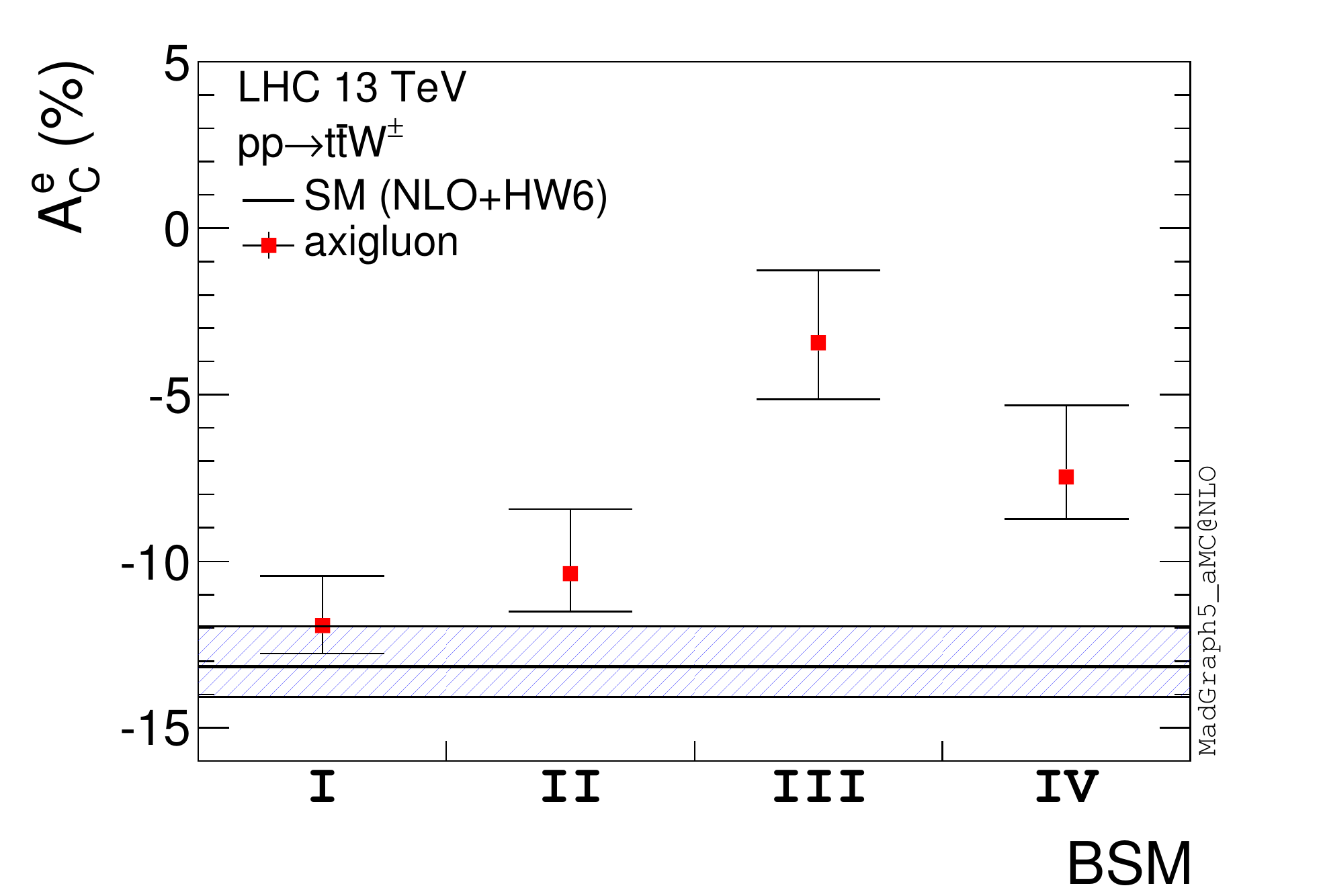}}\\
\caption{Comparison between asymmetries predicted in the axigluon (NLO+BSM, $I-IV$)  scenarios and the NLO SM prediction at 13 TeV, including scale  uncertainties.}
\label{fig:Total_Effect_13}
\end{figure}

\begin{figure}[t]
\centering
\subfigure{\includegraphics[width= 0.5\textwidth]{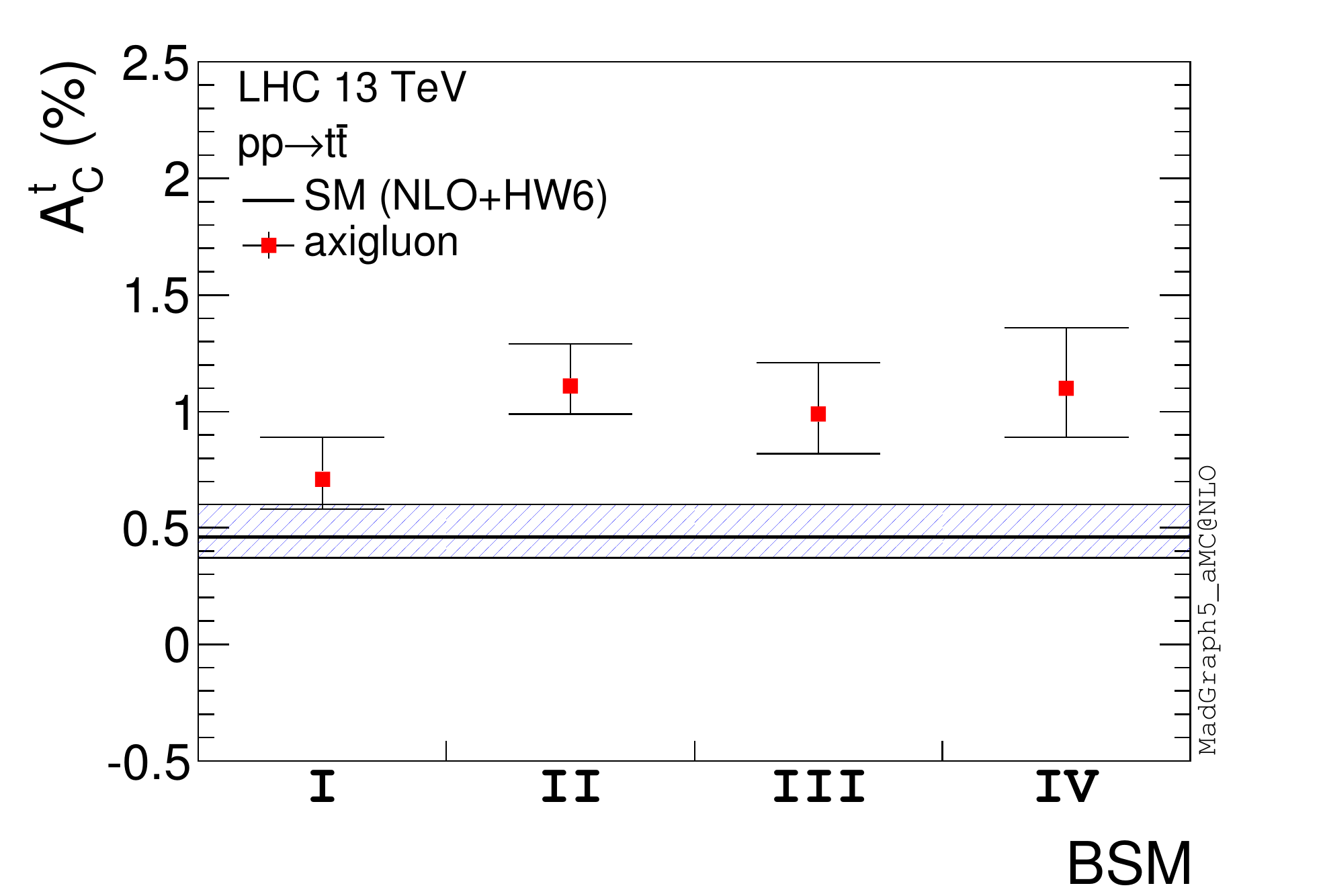}}
\subfigure{\includegraphics[width= 0.5\textwidth]{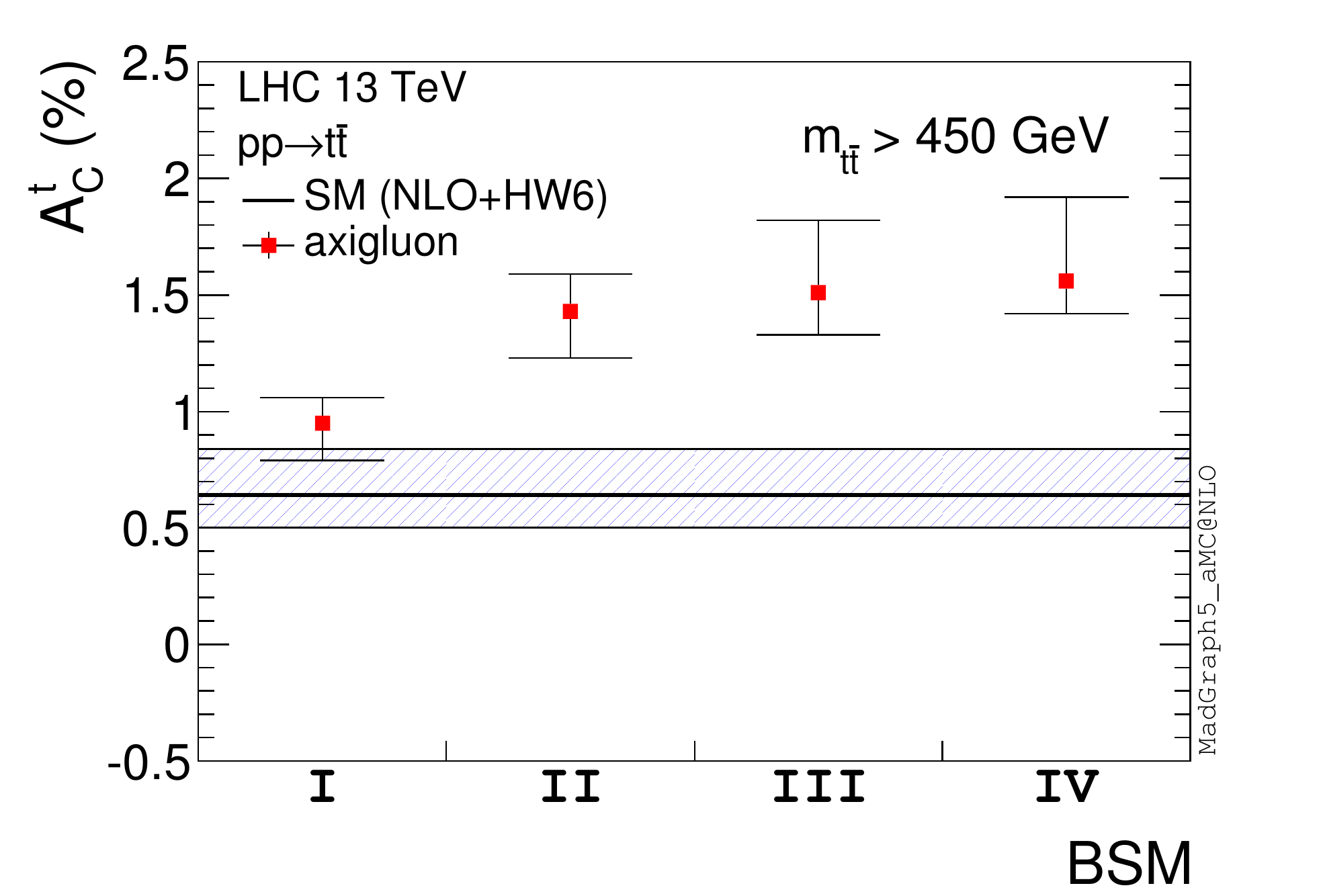}}
\\
\caption{Comparison between asymmetries predicted by axigluon
  (NLO+BSM, $I-IV$) scenarios and the NLO SM prediction at 13 TeV, for
  $t\bar{t}$ production, including scale  uncertainties. Top panel:
  inclusive production. Bottom panel: $m_{t \bar t} > 450$ GeV.}
\label{fig:tt}
\end{figure}

The total asymmetry reads\\
\begin{align}
A_c &= \frac{\sigma^{\rm SM}_{NLO}}{\sigma_{\rm tot}} A^{\rm SM}_c + \frac{\sigma^{\rm BSM}_{\rm LO} }{\sigma_{\rm tot}} A^{\rm BSM}_c  \,.
\label{eq:Effect}
\end{align}

\begin{center}
\begin{table}[ht]
\renewcommand{\arraystretch}{1.5}
\small
\begin{tabular}{ c | c c c }
\hline\hline
 NLO+BSM $(\mu_f = \mu_r = 2m_t)$ & $A_c^{t}(\%)$ & $A_c^{b}(\%)$ & $A_c^{\ell}(\%)$ \\ 
 [0.5ex] 
\hline 
$m_{\tilde G} = 200 \; \text{GeV}$, left-handed & $5.23  $ & $10.67  $ & $-13.42  $ \\ 
 [0.5ex] 
\hline 
$m_{\tilde G}  = 200 \; \text{GeV}$, axial  & $6.69 $ & $11.55  $ & $-11.96  $ \\ 
 [0.5ex] 
\hline
$m_{\tilde G}  = 2000 \; \text{GeV}$, left-handed & $8.76  $ & $13.50 $ & $-9.02 $ \\
 [0.5ex] 
\hline 
 $m_{\tilde G}  = 2000 \; \text{GeV}$, axial  & $7.63 $ & $12.55  $ & $-10.25 $ \\
\hline
\end{tabular}
\caption{Total asymmetries $A_c^{i}$, calculated for $p p \rightarrow t \bar t W^{\pm}$ at $8 \; \textrm{TeV}$. Figures in the table have around 0.1 (\%) of statistical uncertainty.}
\label{table:NLO_BSM_8}
\end{table}
\end{center}
\begin{center}
\begin{table}[h]
\renewcommand{\arraystretch}{1.5}
\small
\begin{tabular}{ c | c c c }
\hline\hline
 NLO+BSM $(\mu_f = \mu_r = 2m_t)$ & $A_c^{t}(\%)$ & $A_c^{b}(\%)$ & $A_c^{\ell}(\%)$ \\ 
 [0.5ex] 
\hline 
$m_{\tilde G} = 200 \; \text{GeV}$, left-handed & $4.73  $ & $9.91  $ & $-11.93  $ \\ 
 [0.5ex] 
\hline 
$m_{\tilde G}  = 200 \; \text{GeV}$, axial  & $6.28 $ & $10.61  $ & $-10.37  $ \\ 
 [0.5ex] 
\hline
$m_{\tilde G}  = 2000 \; \text{GeV}$, left-handed & $11.54  $ & $15.53 $ & $-3.45 $ \\
 [0.5ex] 
\hline 
 $m_{\tilde G}  = 2000 \; \text{GeV}$, axial  & $7.35 $ & $11.13  $ & $-7.46 $ \\
\hline
\end{tabular}
\caption{Total asymmetries $A_c^{i}$, calculated for $p p \rightarrow t \bar t W^{\pm}$ at $13 \; \textrm{TeV}$. Figures in the table have around 0.1 (\%) of statistical uncertainty.}
\label{table:NLO_BSM}
\end{table}
\end{center}

\noindent
All results include showering and hadronization. The effect of the
axigluon BSM is calculated using Eq.~(\ref{eq:Effect}) and the results
are presented in Tables~\ref{table:NLO_BSM_8} and \ref{table:NLO_BSM} (at 8 and 13 TeV respectively),  with the scales $\mu_f =
\mu_r = 2m_t$. We include the uncertainties due to scale variation
($\mu_f = \mu_r = m_t , 2m_t, 4m_t$) separately for the three $A_c^i$
asymmetries in Fig.~\ref{fig:Total_Effect_8}
(Fig.~\ref{fig:Total_Effect_13}) for $\sqrt{s}=8$~TeV
($\sqrt{s}=13$~TeV). 

To compare with the sensitivity of the standard charge asymmetry in
$t\bar{t}$ final states, we
show the results for 13~TeV in Fig.~\ref{fig:tt}. These plots,
compared to those in Fig.~\ref{fig:Total_Effect_13},  show
that the relative impact of BSM modifications is larger for
the $t\bar{t}W^\pm$ asymmetries than for the $t\bar{t}$ ones. The reason
is that any asymmetry in the $t\bar{t}$ final state, whether induced
by QCD effects or by BSM physics, is largely washed out by the
symmetric contribution due to the $gg$ initial state. Of course the
ultimate reach of measurements in the $t\bar{t}W^\pm$ is challenged by the
reduced statistics; we shall show in the next section that the high
luminosities expected in future runs of the LHC are sufficient to
precisely measure the SM asymmetries, and to expose possible BSM
contributions. 

\section{Outlook and conclusions}
\label{sec:concl}
In the previous sections we have argued that the polarization and
asymmetry effects in $t \bar t W^{\pm}$ production are large enough to
offer a useful handle to constrain new physics effects. The question,
however, is whether such effects will be measurable given the expected
cross sections and luminosities at present and future colliders. 
To this aim, we have calculated the cross section for the $t \bar t
W^{\pm}$ process (Tab.~\ref{table:xsec}) at various $pp$ collider
energies, as well as the corresponding  
asymmetries $A_c^i , i = t,b,e$. For comparison, we also show
the results for inclusive $t \bar t$ production. 

To start with, we observe the steady reduction with beam energy 
of the leptonic and $b$
asymmetries. This is due to the growing role of the $qg$ initial-state
channel (shown in Tab.~\ref{table:qg}), which dilutes the EW
component of the asymmetry of the decay products. 
The intrinsic QCD component of the asymmetry is nevertheless more stable,
with $A_c^t$ being reduced by at most 20\% over the range 8--100~TeV. 
The values of $A_c^{b,\ell}$ obtained at 100~TeV by suppressing the
spin correlations are $A_c^{b} = 1.47$ and $A_c^{\ell} = 1.55$, once
again close to the value of $A_c^{t} = 1.85$.   

The charge asymmetry in $t \bar t$ production, viceversa, is reduced
by a factor $\sim 6$ when increasing the energy from 8 to
100~TeV. This is due to two effects related to the small-$x$ behavior
of the PDF's: first the $gg$ channel, which is symmetric and therefore
enters only in the denominator of Eq.~(\ref{eq:A_c}) becomes more and
more dominant; second the $q$ and $\bar q$ asymmetry at large
rapidities is less and less pronounced.

\begin{table*}[ht]
\renewcommand{\arraystretch}{1.5}
\begin{center}
\small
\begin{tabular}{  c | c | c c c c c  }
\hline\hline
 & & 8 TeV &  13 TeV & 14 TeV   & 33 TeV & 100 TeV\\
\hline
 \multirow{2}{*}{$ t \bar t $} &  $ \sigma $(pb)  & 
  $198^{+15\%}_{-14\%}{}$ & 
 $661^{+15\%}_{-13\%}{}$ &
  $786^{+14\%}_{-13\%}{}$ & 
  $4630^{+12\%}_{-11\%}{}$ & 
  $30700^{+13\%}_{-13\%}{}$ \\
  \cline{2-7} 
 & $ A_c^t (\%) $  &  $0.72^{+0.14}_{-0.09} $  & $0.45^{+0.09}_{-0.06} $ & $0.43^{+0.08}_{-0.05} $ & $0.26^{+0.04}_{-0.03} $ & $0.12^{+0.03}_{-0.02}$ \\
\hline 
 \multirow{4}{*}{$ t \bar t  W^\pm $} &  $ \sigma $(fb)  & 
 $210^{+11\%}_{-11\%}{}$ & 
 $587^{+13\%}_{-12\%}{}$ &
  $678^{+14\%}_{-12\%}{}$ & 
  $3220^{+17\%}_{-13\%}{}$ & 
  $19000^{+20\%}_{-17\%}{}$ \\
  \cline{2-7}  
 & $ A_c^t (\%)$  & $2.37^{+0.56}_{-0.38} $ & $2.24^{+0.43}_{-0.32} $ & $2.23^{+0.43}_{-0.33}$ & $1.95^{+0.28}_{-0.23}$ & $1.85^{+0.21}_{-0.17}$ \\
  \cline{2-7} 
& $ A_c^b (\%)$  & $8.50^{+0.15}_{-0.10} $ & $7.54^{+0.19}_{-0.17}$ & $7.50^{+0.24}_{-0.22}$ & $5.37^{+0.22}_{-0.30}$ & $3.36^{+0.15}_{-0.19}$ \\
& $ A_c^e (\%)$  & $-14.83^{-0.65}_{+0.95} $ & $-13.16^{-0.81}_{+1.12}$ & $-12.84^{-0.81}_{+1.11}$ & $-9.21^{-0.87}_{+1.05}$ & $-4.94^{-0.63}_{+0.72}$ \\
\hline
\end{tabular}
\caption{NLO+PS cross sections for $t \bar t$ and $t \bar t W^{\pm}$
  and corresponding asymmetries at several cms energies. The quoted
  uncertainties are estimated with scale variations.}
\label{table:xsec}
\end{center}
\end{table*}

\begin{table*}[t]
\renewcommand{\arraystretch}{1.5}
\begin{center}
\small
\begin{tabular}{  c | c c c c c  }
\hline\hline
  &  8 TeV &  13 TeV & 14 TeV   & 33 TeV & 100 TeV\\
\hline
$t \bar t W^+$, ($qg,\bar q g$) (\%) & 7.5 & 15 & 17 & 33 & 51\\
\hline
\end{tabular}
\caption{Contribution of the $qg$ parton subprocess at NLO for the $t \bar t W^+$ process for $\mu_f = \mu_r = 2m_t$.}
\label{table:qg}
\end{center}
\end{table*}

To derive a quantitative estimate of the statistical precision that
could be optimistically reached under various energies and luminosity
scenarios, we assume leptonic ($\ell = e, \mu$) decays for the top quarks
\begin{align*}
\sigma = \sigma (t \bar t W^{\pm})  \cdot {\rm BR}(t \rightarrow b l^+ \nu_l) ^2 =  0.0484 \cdot  \sigma (t \bar t W^{\pm}) \;,
\end{align*}
\noindent
and neglect acceptance and reconstruction efficiencies.  Using the
results collected in Tab.~\ref{table:xsec} we find:

\begin{itemize}
\item \; 8 TeV ($\mathcal L = 40 \; \text{fb}^{-1}$):\\[5pt]
 $ \delta_{\text{rel}} A_c^t = 209 \% ,  \delta_{\text{rel}} A_c^b = 58 \% ,  \delta_{\text{rel}} A_c^{\ell} = 33 \% $
\item  \; 14 TeV ($\mathcal L = 300 \; \text{fb}^{-1}$):\\[5pt]
 $\delta_{\text{rel}} A_c^t = 45 \% ,  \delta_{\text{rel}} A_c^b = 13 \% ,  \delta_{\text{rel}} A_c^{\ell} = 8 \% $
 \item  \; 14 TeV ($\mathcal L = 3000 \; \text{fb}^{-1}$):\\[5pt]
 $\delta_{\text{rel}} A_c^t = 14 \% ,  \delta_{\text{rel}} A_c^b = 4 \% ,  \delta_{\text{rel}} A_c^{\ell} = 2 \% $
\item  \; 100 TeV ($\mathcal L = 3000 \; \text{fb}^{-1}$):\\[5pt]
 $\delta_{\text{rel}} A_c^t = 3 \% ,  \delta_{\text{rel}} A_c^b = 2 \% ,  \delta_{\text{rel}} A_c^{\ell} = 1 \% $
\end{itemize}
where $\delta_{rel}A=\delta A/A$ is the relative precision on the
asymmetries. While a realistic experimental analysis will certainly
degrade this optimal precision, these numbers show the great potential
of this observable. 

We remark that the larger sensitivity of $A_c^{b,\ell}$ compared to
$A_c^{t}$ follows from the larger value of the former compared to the
latter. The sensitivity to the purely QCD component of
$A_c^{b,\ell}$,  however, is comparable to the sensitivity of
$A_c^{t}$. For example, at 100~TeV $\delta_{\text{rel}} A_c^{\ell}=1\%$
implies $\delta A_c^{\ell} \sim 0.0005$, which is about 3\% of its
QCD component, a precision consistent with what we quote for $A_c^t$.

In conclusion, the main motivation of our work has been the
observation that the top quark charge asymmetry in  $pp \rightarrow t
\bar t W^{\pm}$ at the LHC is larger than that of inclusive $t \bar
t$, being of  a few percents. In addition,  the lepton and $b$
asymmetries are very large and already present at the leading order
due to the polarization of the initial fermionic line by the $W^{\pm}$
emission. As a simple application, we have shown how the existence of
an axigluon that could describe the Tevatron measurements of the
forward-backward asymmetry would impact  $pp \rightarrow t \bar t
W^{\pm}$ and discussed the prospects in LHC Run II, HL-LHC  and at  future
colliders. 

The $t\bar{t} W^\pm$ final state will not replace the use of  the
$t\bar{t}$ asymmetry, particularly while the total integrated
luminosity of the LHC is still below the O(100~fb$^{-1}$). In the
long term, however, it will provide a powerful probe, complementary to the
$t\bar{t}$ asymmetry, and uniquely sensitive to the chiral nature of
possible new physics that were to manifest itself in these measurements. 

\section{Acknowledgements}
We are thankful to Josh McFayden, Tamara Vazquez Schroeder and Elizaveta Shabalina for 
drawing our attention to features in $ttW^\pm$ MC simulations that inspired this investigation. 
We thank James Wells for useful remarks and discussions. 
This work has been supported in part by the ERC grant 291377 ``LHC Theory'',  
by the Research Executive Agency (REA) of the European Union under the Grant Agreement numbers PIT-GA-2010-264564 (LHCPhenoNet) and PITN-GA-2012-315877 (MCNet). The work of FM and IT is supported by the IISN
``MadGraph'' convention 4.4511.10, by the IISN ``Fundamental interactions''
convention 4.4517.08, and in part by the Belgian Federal Science Policy Office
through the Interuniversity Attraction Pole P7/37. The work of MZ has been 
partly supported by the ILP LABEX (ANR-10-LABX-63), in turn supported by 
French state funds managed by the ANR within the ``Investissements d’Avenir''
programme under reference ANR-11-IDEX-0004-02.


\appendix

\section{$q_L \bar q_R \to t \bar t$ vs $q \bar q \to t \bar tW^\pm$}
\label{sec:basics}

We first review the main features of polarized $q_L \bar q_R \to t \bar t$ scattering, on the same lines as $e^-_L e^+_R \rightarrow t \bar t$ is discussed in Ref.~\cite{Parke:1996pr}. In the beam line basis, {\it i.e.}, when the polarization axis of the top is the light antiquark direction in the top rest frame, the polarized differential cross sections $d\sigma_{ t {\rm pol}, {\bar t} {\rm pol}}$ for an initial state $q_L~\bar q_R$ pair  read
\begin{eqnarray}
&&\mkern-24mu 
    \frac{d\sigma_{\uparrow\uparrow}}{d\cos\theta^*}   = 
     \frac{d\sigma_{\downarrow\downarrow}}{d\cos\theta^*}  =  {\cal N}(\beta) \frac{\beta^2(1-\beta^2) \sin^2 \theta^* }
    {(1+\beta \cos \theta^*)^2} 
    \, , \nonumber \\[0.1in]
&&\mkern-24mu 
    \frac{d\sigma_{\downarrow\uparrow}}{d\cos\theta^*}   = 
    {\cal N}(\beta) ~\frac{\beta^4 \sin^4 \theta^*}{(1+\beta \cos \theta^*)^2} 
    \, , \nonumber \\[0.1in]
&&\mkern-24mu 
    \frac{d\sigma_{\uparrow\downarrow}}{d\cos\theta^*}  = 
    {\cal N}(\beta)
     \frac{[(1+ \beta  \cos \theta^*)^2 + (1-\beta^2) ]^2}{(1+\beta\cos\theta^*)^2} \,,
\label{Beamline}
\end{eqnarray}
where ${\cal N}(\beta)$ is a normalization factor 
\begin{equation}
    {\cal N}(\beta) = \frac{\pi \alpha_S^2 }{9 s} \beta \, ,
\end{equation}
and $\cos\theta^*$ is the polar angle of the top quarks in parton-parton centre-of-mass frame.
This basis is useful  both at threshold, ($\beta \to 0$), where it is clear that only one amplitude, $q_L \bar q_R \rightarrow t_{\uparrow} \bar t_{\downarrow}$ is non-zero, meaning that the top quarks are completely polarized, and  at high energy, ($\beta \to 1$), where it is manifest  that the top anti-top  polarizations are opposite, 
\begin{eqnarray}
&& \frac{d\sigma_{\uparrow\downarrow, \downarrow\uparrow}}{d\cos\theta^*}  \stackrel{\beta \to 1}{=} {\cal N}(1) (1 \pm  \cos \theta^*)^2  \,,
\label{beta1}
\end{eqnarray}
a result which is also valid in the helicity basis~\cite{Parke:1996pr}. Eq.~(\ref{beta1}) predicts the total number of events with opposite top anti-top polarization to be the same far from threshold. The polarization information is transferred to the decay products  angular distributions, and in particular to the leptons that are 100\% correlated with the top-quark spins. 
\begin{figure}[h]
\centering
\subfigure{\includegraphics[width= 0.5\textwidth]{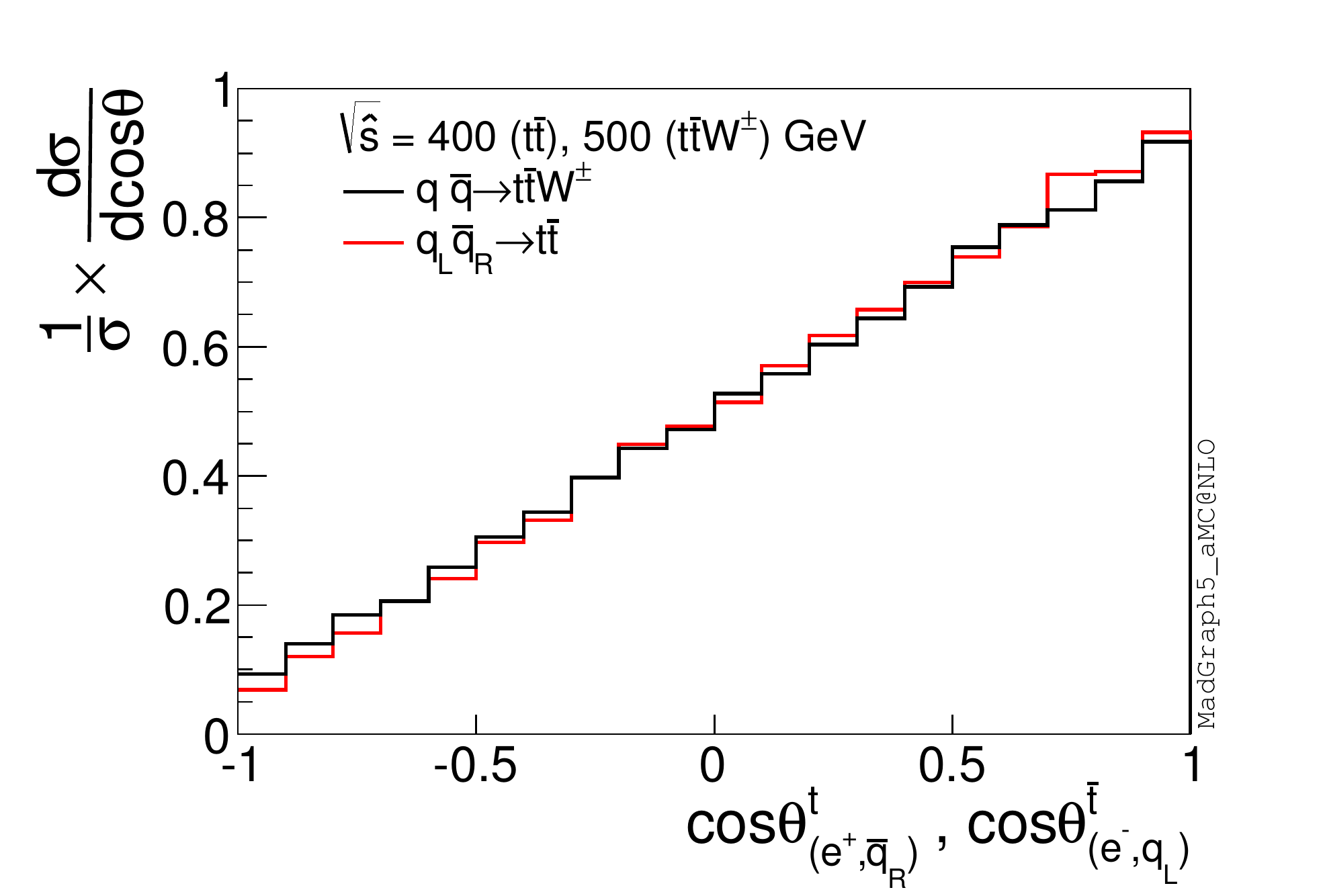}}
\subfigure{\includegraphics[width= 0.5\textwidth]{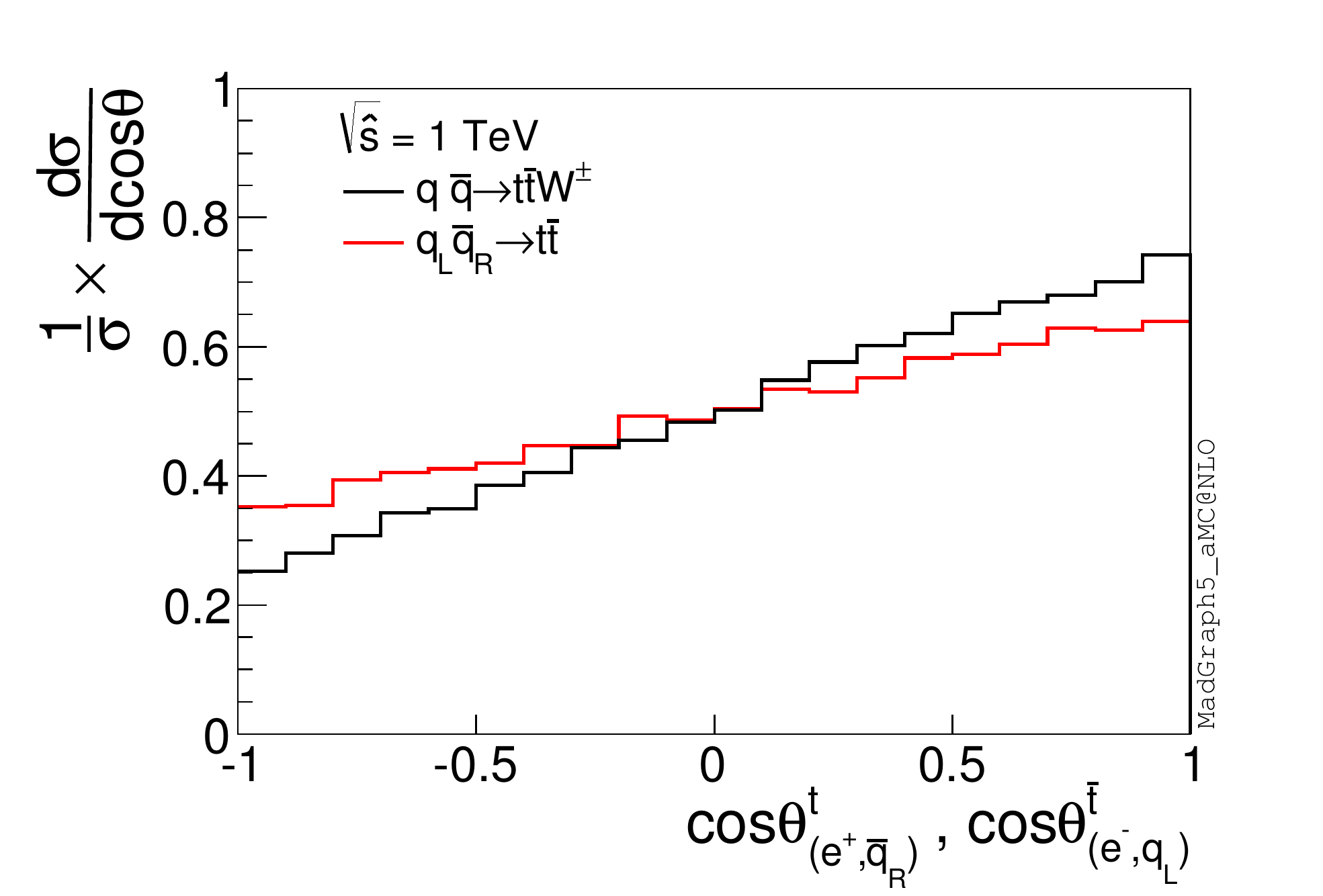}}
\subfigure{\includegraphics[width= 0.5\textwidth]{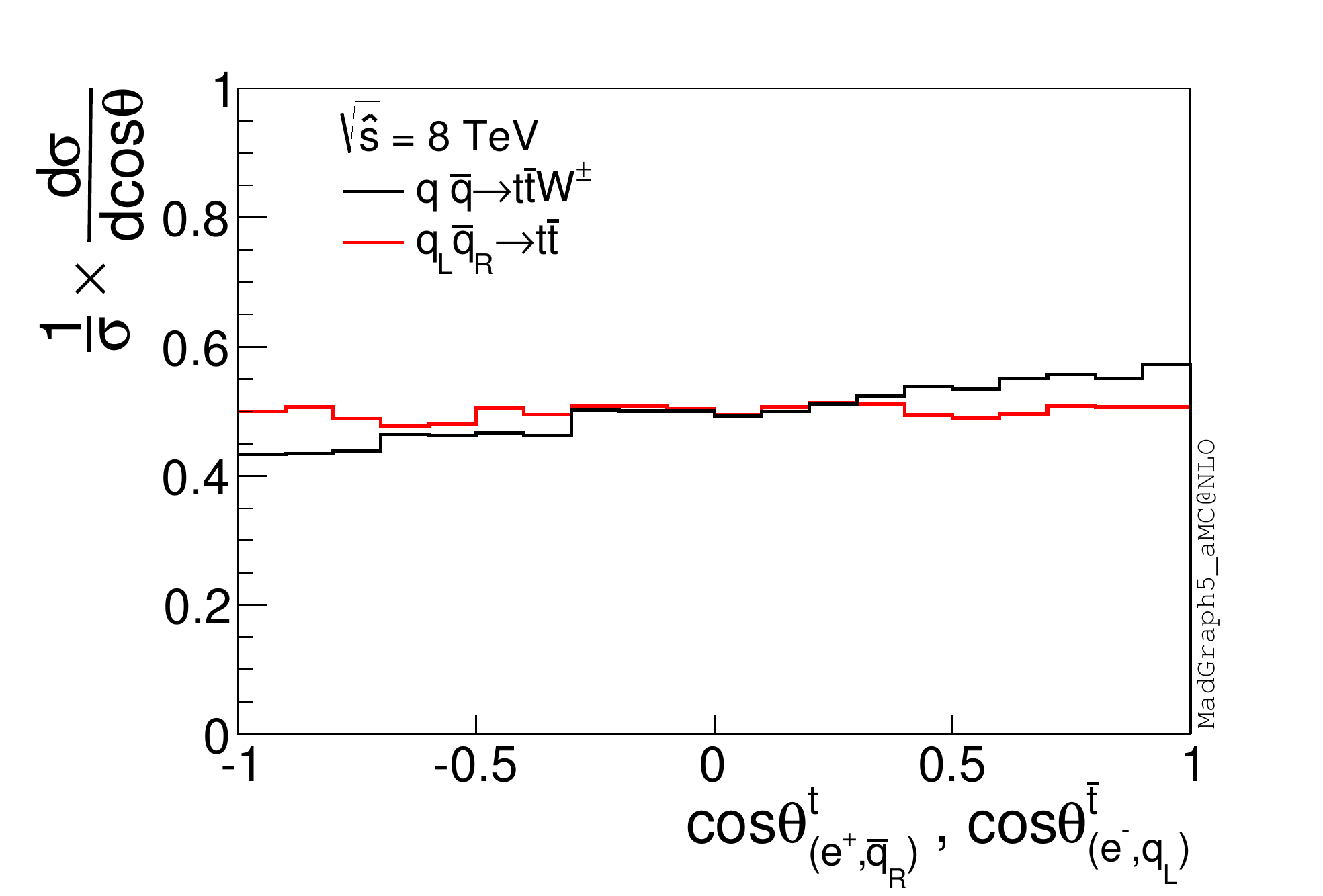}}
\caption{Normalized $\cos{\theta}$  distributions for the leptons with respect to the spin axis for the $t , \bar t$ defined in the beam-axis as in Ref.~\cite{Parke:1996pr} at different parton-parton energies (no PDF's). Close to threshold, {\it i.e.} 400 GeV for $t\bar t$ and  500 GeV for $t\bar t W^\pm$,  the $t$ and  $\bar t$ are fully polarized. As the energy increases the distribution flattens out up to a constant at very high energies in agreement with Eq.~(\ref{beta1}).}
\label{fig:CosthetaPlots}
\end{figure}

\begin{figure}[h]  
\subfigure{\includegraphics[width= 0.5\textwidth]{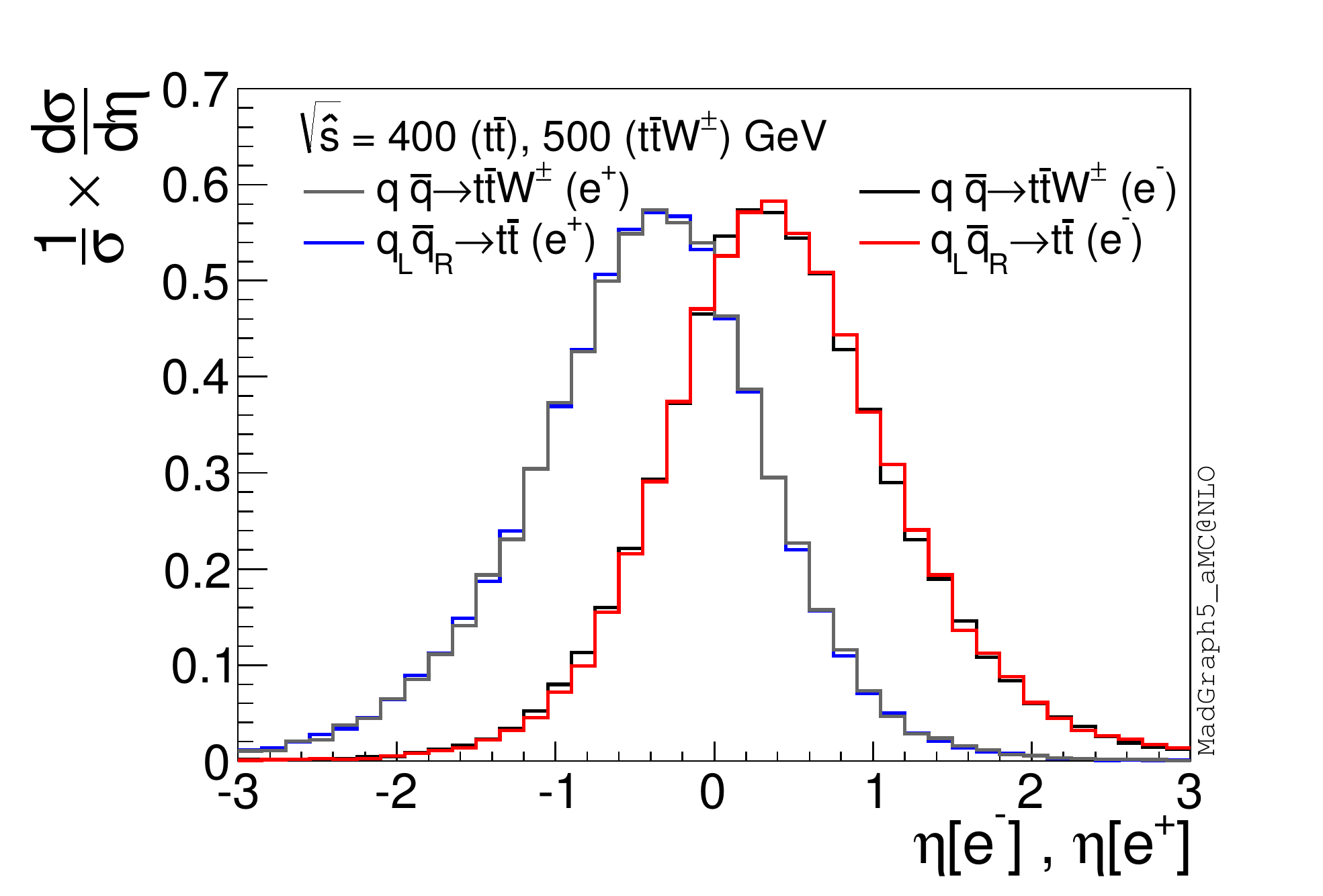}}
\subfigure{\includegraphics[width= 0.5\textwidth]{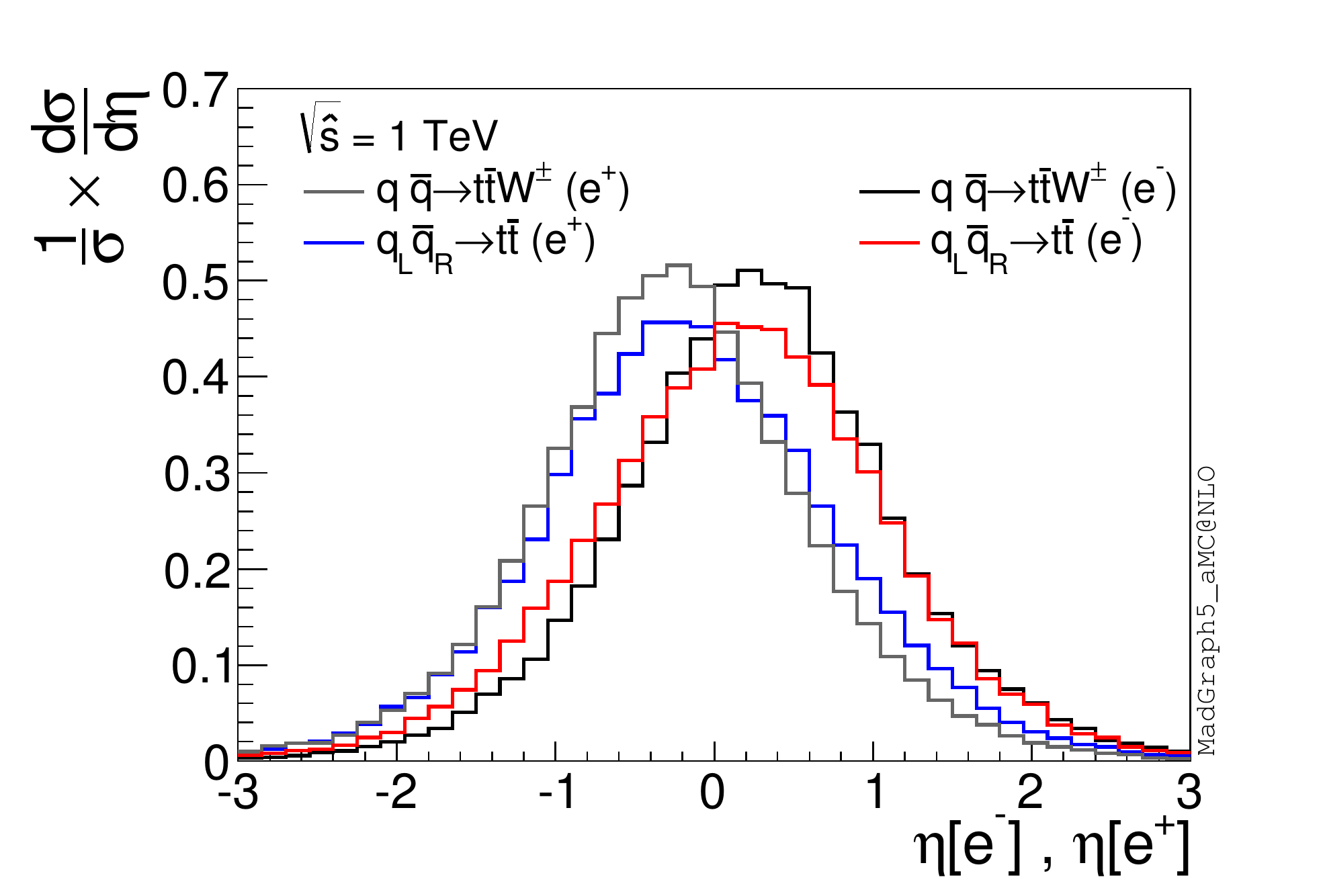}}
\subfigure{\includegraphics[width= 0.5\textwidth]{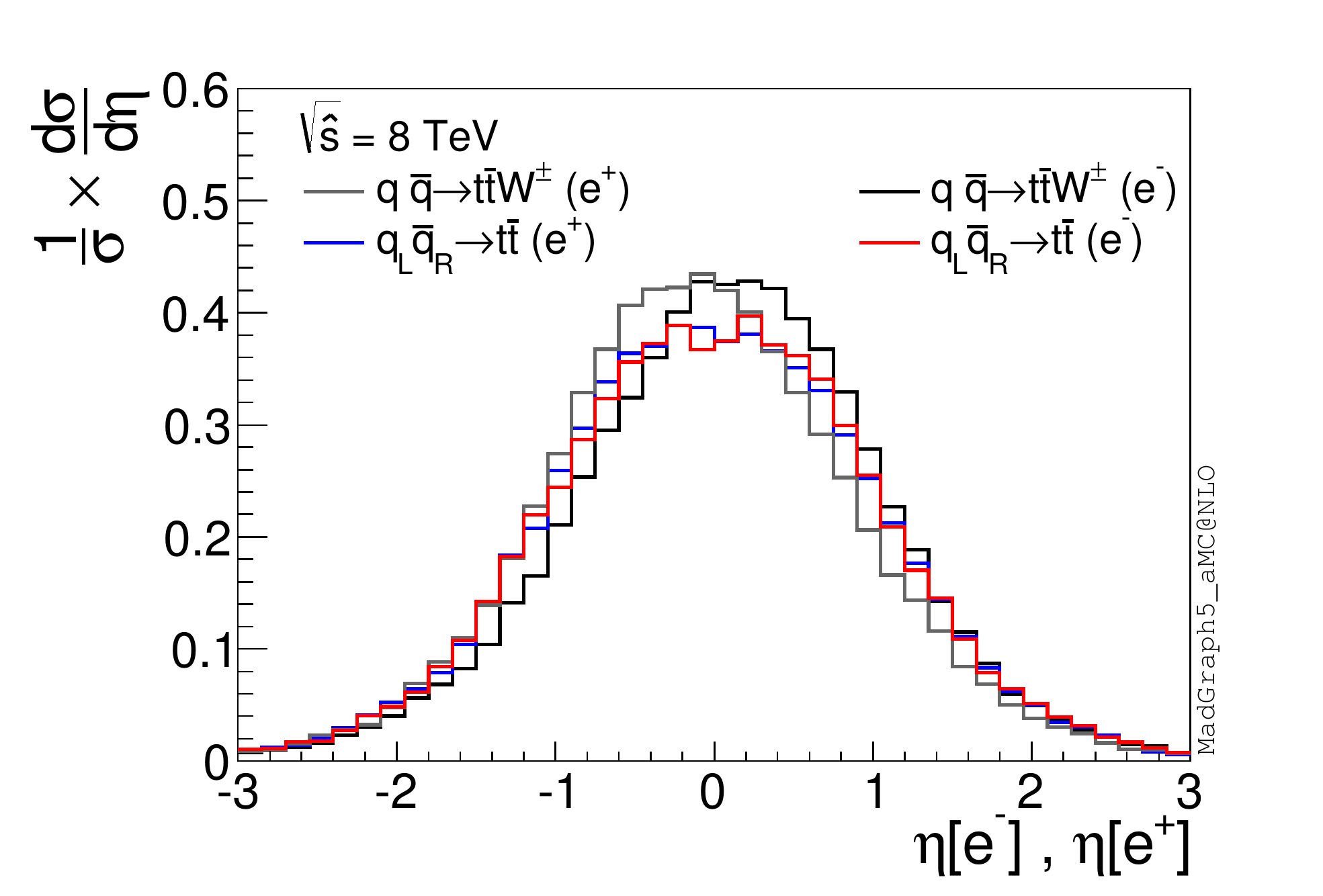}}
\caption{Normalized pseudorapidity distributions in the $t \bar t$ center of mass frame. Close to threshold, {\it i.e.} 400 GeV for $t\bar t$ and  500 GeV for $t\bar t W^\pm$ full polarization of $t$ and  $\bar t$ determines a sizable opposite asymmetry in the distributions of the $e^+$ and $e^-$. Far from threshold, the distribution becomes more and more symmetric.\vspace{2\baselineskip}}
\label{fig:PseudoPlots}
\end{figure}
One therefore expects  the lepton polar distributions with respect to
the beam axis to show a linear dependence in $\cos\theta_e$ at
threshold that flattens out at high energies.  

We have explicitly checked the expressions Eq.~(\ref{Beamline})  and
the analytic computation of the tree-level $q_L \bar q_R \to t\bar t
\to b \ell^+ \nu\,  \bar b \ell^- \nu$ amplitude numerically to those
obtained via {\sc MadGraph5\_aMC@NLO}. Apart
from more complicated analytic formulas the case of $q \bar q \to t
\bar t W^\pm$ is totally analogous, as the only non-trivial effect of
the $W$-boson emission is that of selecting a $q_L \bar q_R$ in the
initial state.  

This is clearly shown  in Figs. \ref{fig:CosthetaPlots} and \ref{fig:PseudoPlots}. 
In the first set of plots we show the lepton distributions from the
top-quarks decay for both $q_L \bar q_R \to t \bar t$ and $q \bar q
\to  t \bar t W^\pm$  in the beam-axis frame at three values of
$\sqrt{\hat s}$, one close to threshold (400 GeV for $t\bar t$ and
500 GeV for $t\bar t W^\pm$)  and increasingly  far from threshold (1
and 8 TeV).  The two processes lead to very similar distributions. We
have then considered the pseudorapidity distributions in the $t \bar t$ rest
frame. We find that the  $t$ and the $\bar t$ pseudorapidity distributions
are equal and symmetric at LO and we do not show them. The lepton
distributions, however, see Fig.~\ref{fig:PseudoPlots}, display an
opposite and equal forward-backward asymmetry whose shapes  in the
centre-of-mass frame of the $t\bar t$ pair  are again extremely
similar in $q_L \bar q_R \to t \bar t$ and $q \bar q \to  t \bar t
W^\pm$. The fact that the asymmetry is larger at threshold is a direct
consequence of the fact that there the top quarks are fully polarized.


\end{document}